\documentclass[letterpaper,journal=jctcce,manuscript=article]{achemso}
\usepackage{geometry}
\makeatletter
\renewcommand{\acs@maketitle@width}{\textwidth}
\makeatother

\usepackage[T1]{fontenc}
\usepackage{libertinus}

\AtBeginDocument{\fontdimen3\font=0.3em}
\usepackage{setspace}\AtBeginDocument{\singlespacing}

\usepackage[colorlinks=true,citecolor=blue,linkcolor=blue,urlcolor=blue]{hyperref}
\usepackage{subcaption}

\newcommand{\torchjit}{torch\allowbreak.jit\allowbreak.script}
\newcommand{\torchcompile}{torch\allowbreak.compile}

\def\InlineFloatEnv{}

\title[OpenMM-Python-Force]
{OpenMM-Python-Force: Deploying Accelerated Python Modules in Molecular Dynamics Simulation}
\author{Zhi Wang}
\affiliation{ByteDance Research, Bellevue, Washington 98004, USA.}
\email{zhi.wang1@bytedance.com}
\author{Wen Yan}
\affiliation{ByteDance Research, Bellevue, Washington 98004, USA.}
\email{wen.yan@bytedance.com}
\date{\today}
 
\makeatletter
\begin{document}

\maketitle

\begin{abstract}

We present OpenMM-Python-Force,
a plugin designed to extend OpenMM's functionality
by enabling integration of energy and force calculations
from external Python programs via a callback mechanism.
During molecular dynamics simulations,
data exchange can be implemented through \texttt{torch.Tensor}
or \texttt{numpy.ndarray}, depending on the specific use case.
This enhancement significantly expands OpenMM's capabilities,
facilitating seamless integration of accelerated Python modules
within molecular dynamics simulations.
This approach represents a general solution that can be adapted
to other molecular dynamics engines beyond OpenMM.
The source code is openly available at
\href{https://github.com/bytedance/OpenMM-Python-Force}{https://github.com/bytedance/OpenMM-Python-Force}.

\end{abstract}
 
\section{Introduction}

Modern computational scientific research is standing at the intersection
of two distinct technical domains:
molecular dynamics (MD) simulation and machine learning (ML).
These fields have evolved along different technological paths,
with MD engines predominantly built on C-family languages
utilizing ahead-of-time (AOT) compilation,
while the ML ecosystem has consolidated around Python,
particularly the PyTorch framework.
This technological divergence creates significant challenges for researchers
seeking to integrate these paradigms effectively.

The incorporation of pre-trained ML models into MD simulations
has primarily relied on \texttt{\torchjit},
which generates TorchScript graphs through static analysis
and Python Abstract Syntax Tree parsing.
While this approach enables just-in-time (JIT) optimizations
and provides essential C++ APIs\textemdash{}exemplified by implementations
like OpenMM Torch\cite{Eastman2024,Software-OpenMM-Torch}\textemdash{}it imposes considerable limitations
by supporting only a restricted subset of Python syntax.
These constraints are substantial in practice,
with approximately 50\% of real-world models failing to compile successfully
using \texttt{\torchjit} \cite{Ansel2024}.

Recent advances have introduced various strategies
to enhance both training and inference performance of ML models.
CUDA 12's Graph functionality enables the recording
and efficient replay of CUDA kernel sequences
to reduce kernel launch overheads.
For cases with well-identified computational bottlenecks,
hand-optimized CUDA operators remain an effective optimization strategy.
Projects like NNPOps \cite{Eastman2024,Software-NNPOps} demonstrate this approach
by providing specialized operators for tasks
such as Particle Mesh Ewald calculations and neighbor list construction.
PyTorch 2.0's \texttt{\torchcompile} \cite{Ansel2024} marks a significant breakthrough,
particularly valuable for non-obvious performance bottlenecks.
The latter employs a frontend that extracts PyTorch operations
from Python bytecode to construct FX graphs,
which are then processed by a compiler backend
generating Triton code for CUDA execution.
This approach substantially relaxes Python syntax restrictions
while enabling sophisticated optimizations such as kernel fusion.
Moreover, it can modify the bytecode to incorporate AOT-compiled kernels.
The efficacy of these optimizations is evidenced
by frameworks such as TorchMD-NET \cite{Pelaez2024},
which has achieved significant performance improvements
in their training pipeline.

Despite its promising capabilities,
\texttt{\torchcompile} has not yet achieved widespread adoption
in molecular dynamics simulations,
primarily due to its lack of native C++ support.
This paper addresses this limitation through three steps.
First, we present a general callback mechanism that enables any Python module
to serve as a gradient provider for MD simulations.
Second, we validate this approach through rigorous numerical testing and
performance profiling in gas-phase simulations,
demonstrating both its numerical accuracy and computational efficiency.
Finally, we demonstrate the broad applicability of this approach
through its implementation in ab initio molecular dynamics (AIMD) simulations
and explore its portability to other MD engines,
illustrating how our solution provides a generic and seamless integration
between Python modules and MD simulations.
 
\section{Methods}

The fundamental basis for the callback mechanism lies
in the C-API of the CPython interpreter. When Python code such as

\footnotesize\begin{verbatim}
results = model(*args, **kwargs)
\end{verbatim}\normalsize
is executed, it is internally translated into an equivalent C function call

\footnotesize\begin{verbatim}
PyObject *results, *model, *args, *kwargs;
PyObject *id = model;
results = PyObject_Call(id, args, kwargs);
\end{verbatim}\normalsize
This direct mapping between Python code and C function calls,
as illustrated in Figure~\ref{fig:Callback-C-API-vs-pybind11},
extends to all Python operations,
making it theoretically feasible to translate any Python code sequence
into equivalent C function calls.
While such manual translation would be impractical,
the pybind11 library \cite{Software-pybind11} substantially simplifies this process.
The implementation of the callback mechanism requires two key components:
(1) obtaining the unique identifier (a long integer) of the callable object in
Python,
which corresponds to the value of its \verb|PyObject| pointer,
achieved using Python's built-in \verb|id()| function, and
(2) transferring and storing this identifier from Python to C,
accomplished through the CPython API or binding libraries
such as pybind11 or SWIG \cite{Software-SWIG}.

After capturing the \verb|PyObject| pointer of the callable Python object
in C and ensuring that the object remains resident in memory,
model inference can proceed independently
of the ML model's deployment method or optimization strategy.
This approach establishes a robust and flexible foundation
for integrating Python-based ML models with C-based MD simulations.

\ifdefined\InlineFloatEnv
\begin{figure*}[htbp]
\includegraphics[width=\textwidth]{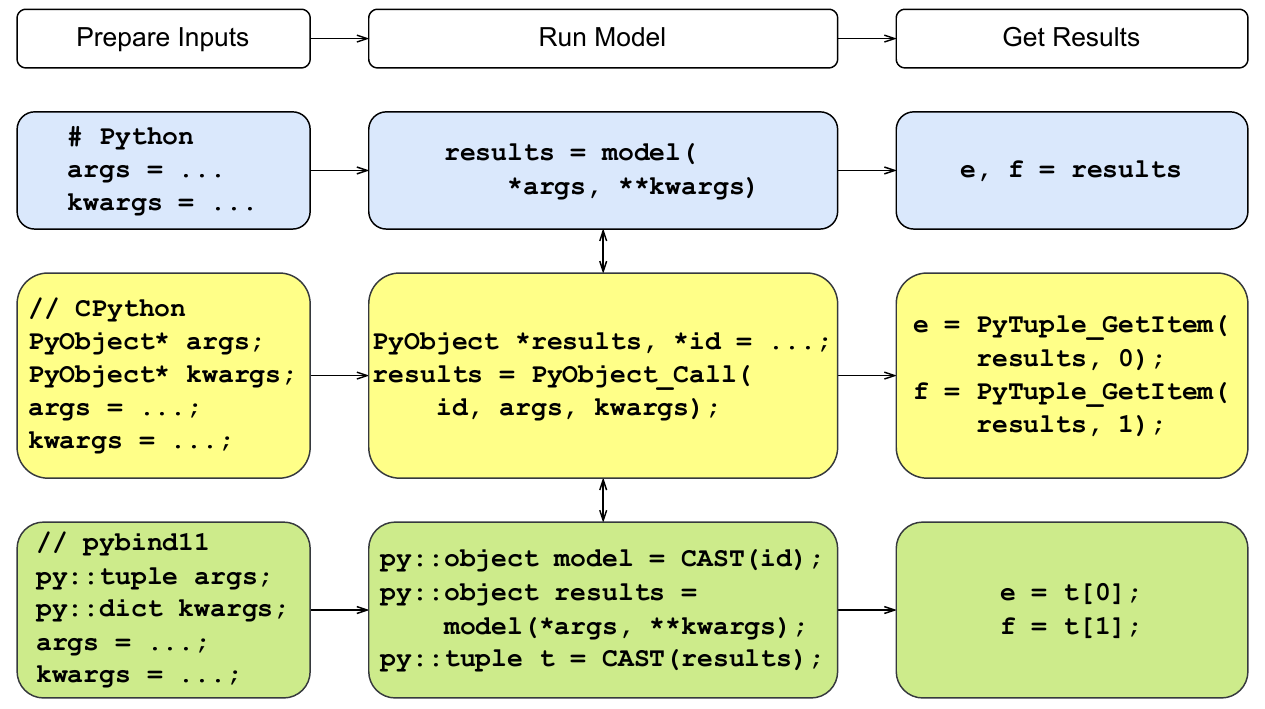}
\caption{Illustration of the Python callback mechanism,
demonstrating the translation between a Python function call
and its corresponding pseudo C/C++ implementations
using either the CPython API or pybind11
(with the C++ namespace \texttt{pybind11} abbreviated
as \texttt{py}).}\label{fig:Callback-C-API-vs-pybind11}
\end{figure*}
 \else\fi

The implementation is streamlined through a custom \verb|Callable| class
that encapsulates the model identifier, the return variables,
and the function parameters.
This class serves as the primary initialization parameter
for the new OpenMM force provided in the plugin.
Integration of this mechanism into an existing OpenMM Python script
is straightforward, as demonstrated below:

\footnotesize\begin{verbatim}
from CallbackPyForce import Callable, TorchForce
class Model42(torch.nn.Module):
    def forward(self, positions):
        return torch.sum(positions**2)
model42 = Model42()
model42 = torch.compile(model42)
call = Callable(id(model42), Callable.RETURN_ENERGY)
force = TorchForce(call)
openmm_system.addForce(force)
\end{verbatim}\normalsize
The Torch library handles force calculations through backpropagation
when the forward pass does not explicitly compute forces.
Additionally, this mechanism seamlessly supports model optimization
through PyTorch's compilation tools:
models can be enhanced with \texttt{\torchjit} or \texttt{\torchcompile}
simply by adding these statements to the existing code,
requiring minimal additional modifications.
 
\section{Results and Discussion}

\subsection{Example: Ethanol}

We evaluated the numerical accuracy and computational performance
using a single ethanol molecule in vacuum,
implementing eight distinct deployment
and compilation strategies (Table~\ref{tb:deploy8}).
For each strategy, we performed independent NVE simulations
for 100 steps with a 1~fs time-step.
All simulations utilized identical random seeds
and initial velocities corresponding to 300~K.
We employed the BAMBOO \cite{Gong2024} MLFF throughout,
configuring OpenMM simulations with the mixed-precision CUDA platform
and setting the BAMBOO model's internal data type to fp32.
All computations were executed on a single NVIDIA L4 GPU.
The numerical accuracy analysis encompassed three evaluations.

\ifdefined\InlineFloatEnv
\begin{table*}[htbp]
\caption{Deployment and compilation strategies evaluated
in ethanol simulations.}\label{tb:deploy8}
\begin{tabular}{c|cr|cr}
\hline
       & No. & Description                   & No. & Description   \\\hline
C++    & 1   & OpenMM Torch (Baseline)       & 5  & 1 + CUDA Graph \\
Python & 2   & native \verb|torch.nn.Module| & 6  & 2 + CUDA Graph \\
Python & 3   & \texttt{\torchjit}            & 7  & 3 + CUDA Graph \\
Python & 4   & \texttt{\torchcompile}        & 8  & 4 + CUDA Graph \\\hline
\end{tabular}
\end{table*}
 \else\fi

\textbf{Hamiltonian Conservation}
The time evolution of the Hamiltonian,
shown in Figure~\ref{fig:drift}, exhibited excellent consistency
in all eight simulations.
The system maintained an average total energy of $-2545.1$~kJ/mol
with a small standard deviation of $0.22$~kJ/mol.

\ifdefined\InlineFloatEnv
\begin{figure*}[htbp]
\begin{subfigure}{0.48\textwidth}
\includegraphics[width=\linewidth]{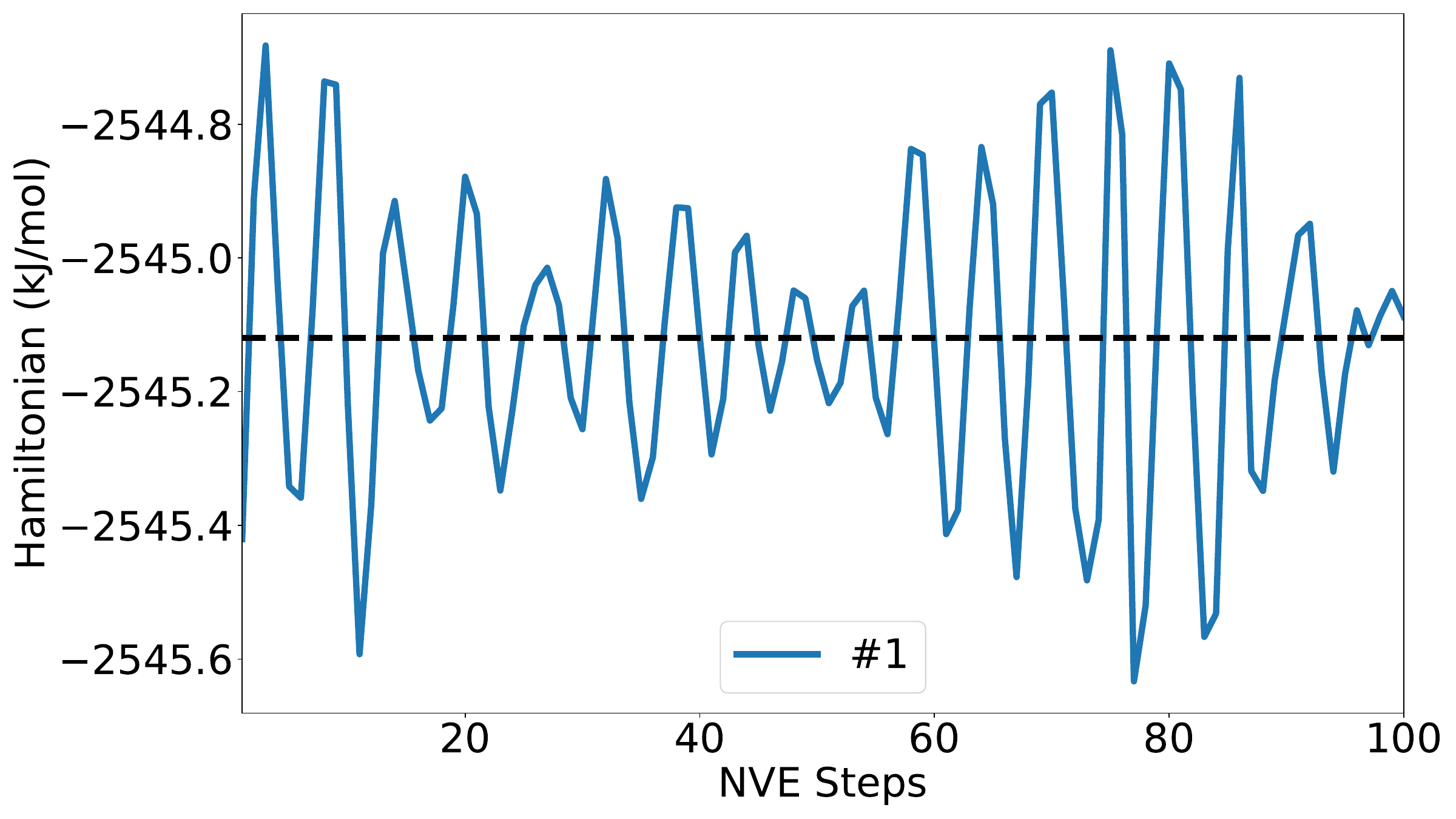}
\end{subfigure}
\begin{subfigure}{0.48\textwidth}
\includegraphics[width=\linewidth]{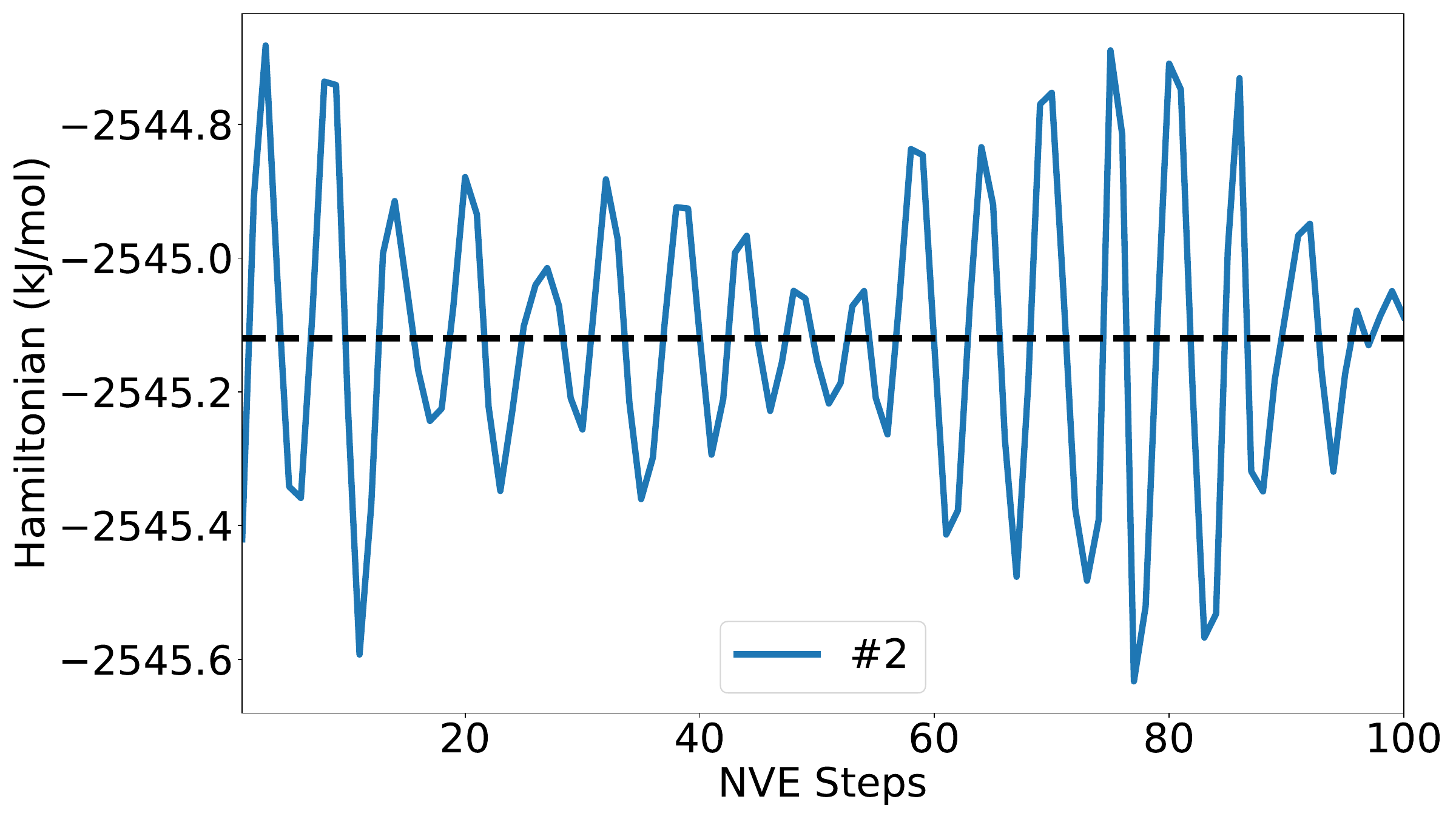}
\end{subfigure}
\\
\begin{subfigure}{0.48\textwidth}
\includegraphics[width=\linewidth]{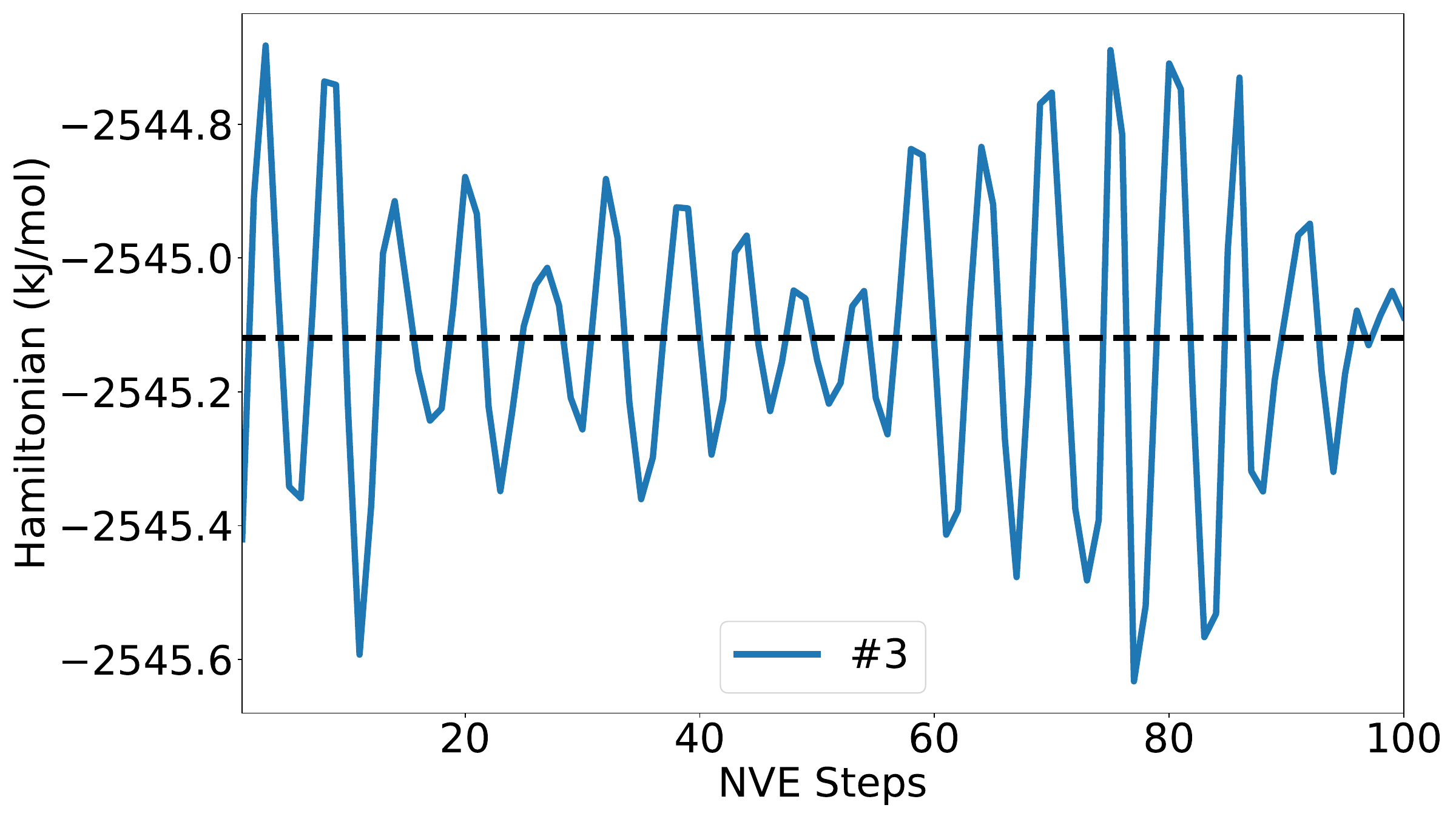}
\end{subfigure}
\begin{subfigure}{0.48\textwidth}
\includegraphics[width=\linewidth]{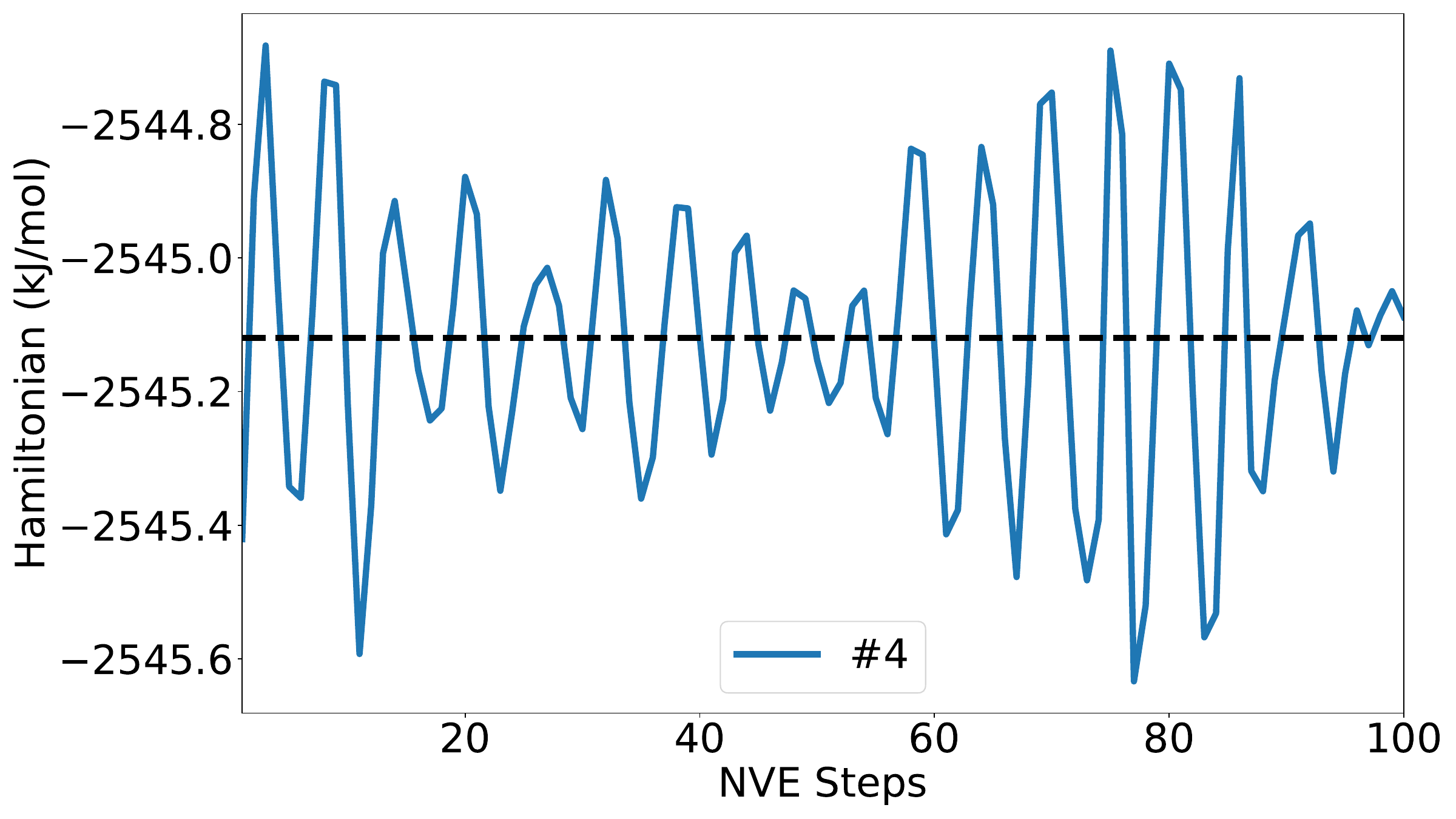}
\end{subfigure}
\\
\begin{subfigure}{0.48\textwidth}
\includegraphics[width=\linewidth]{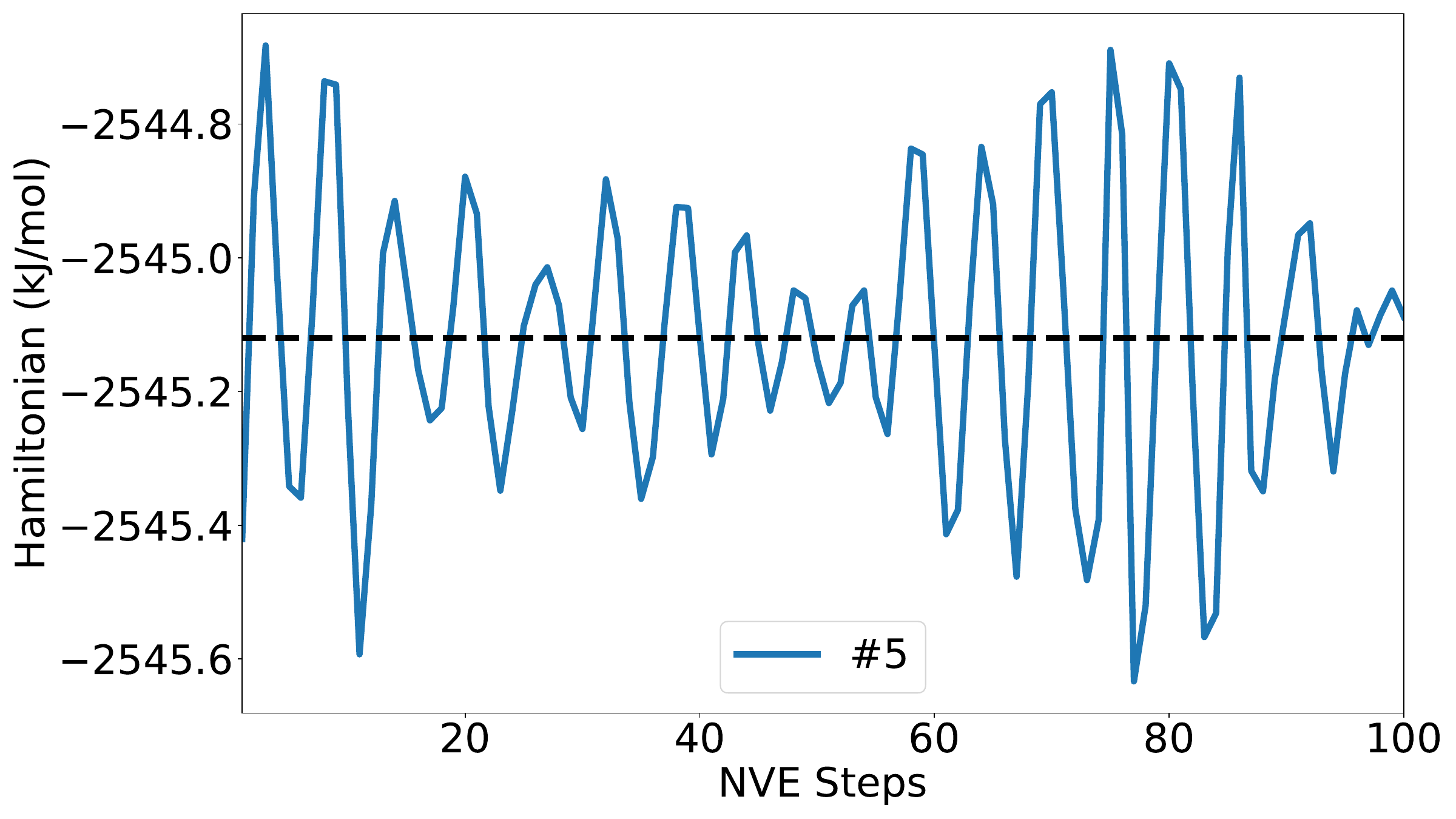}
\end{subfigure}
\begin{subfigure}{0.48\textwidth}
\includegraphics[width=\linewidth]{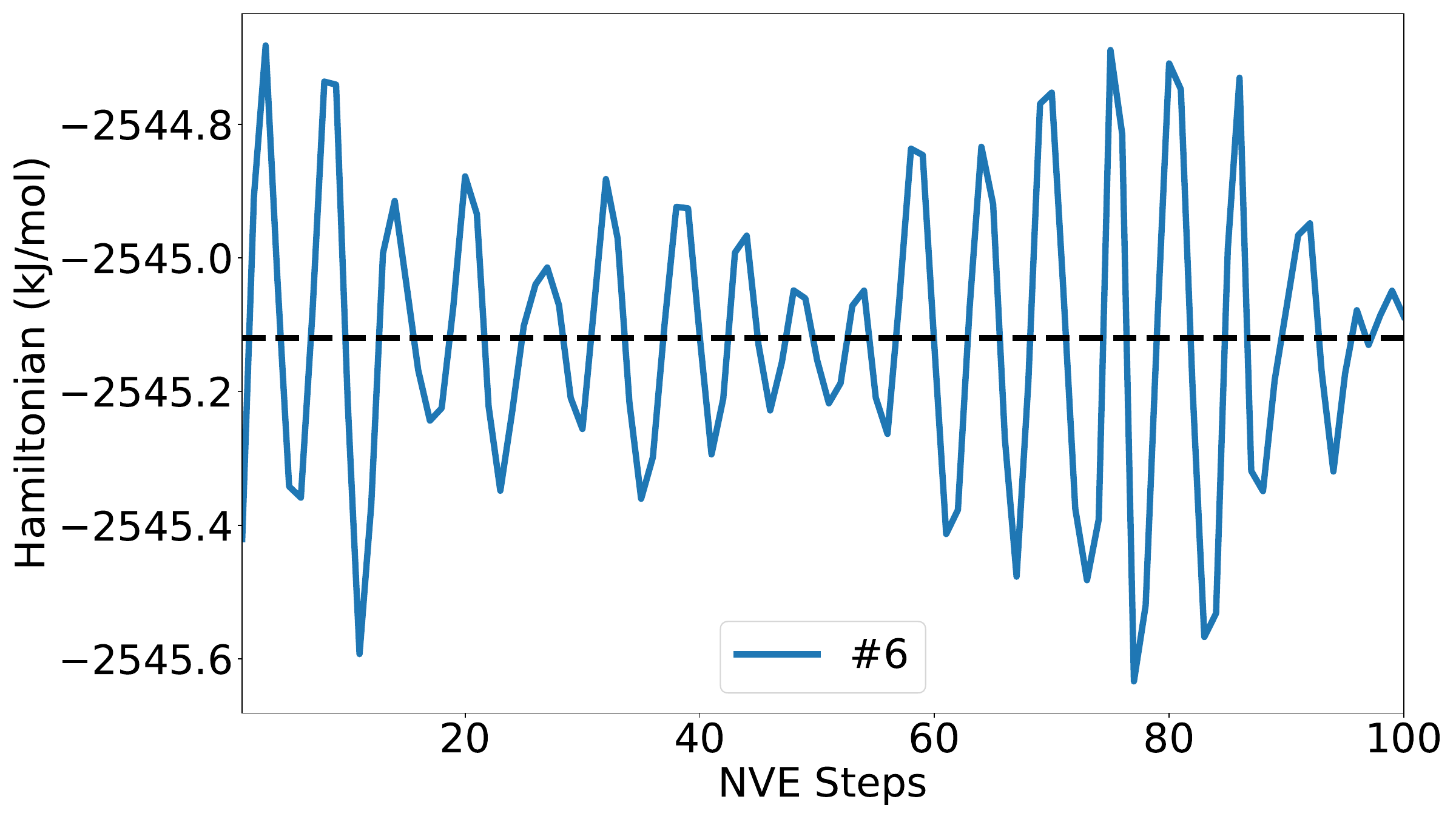}
\end{subfigure}
\\
\begin{subfigure}{0.48\textwidth}
\includegraphics[width=\linewidth]{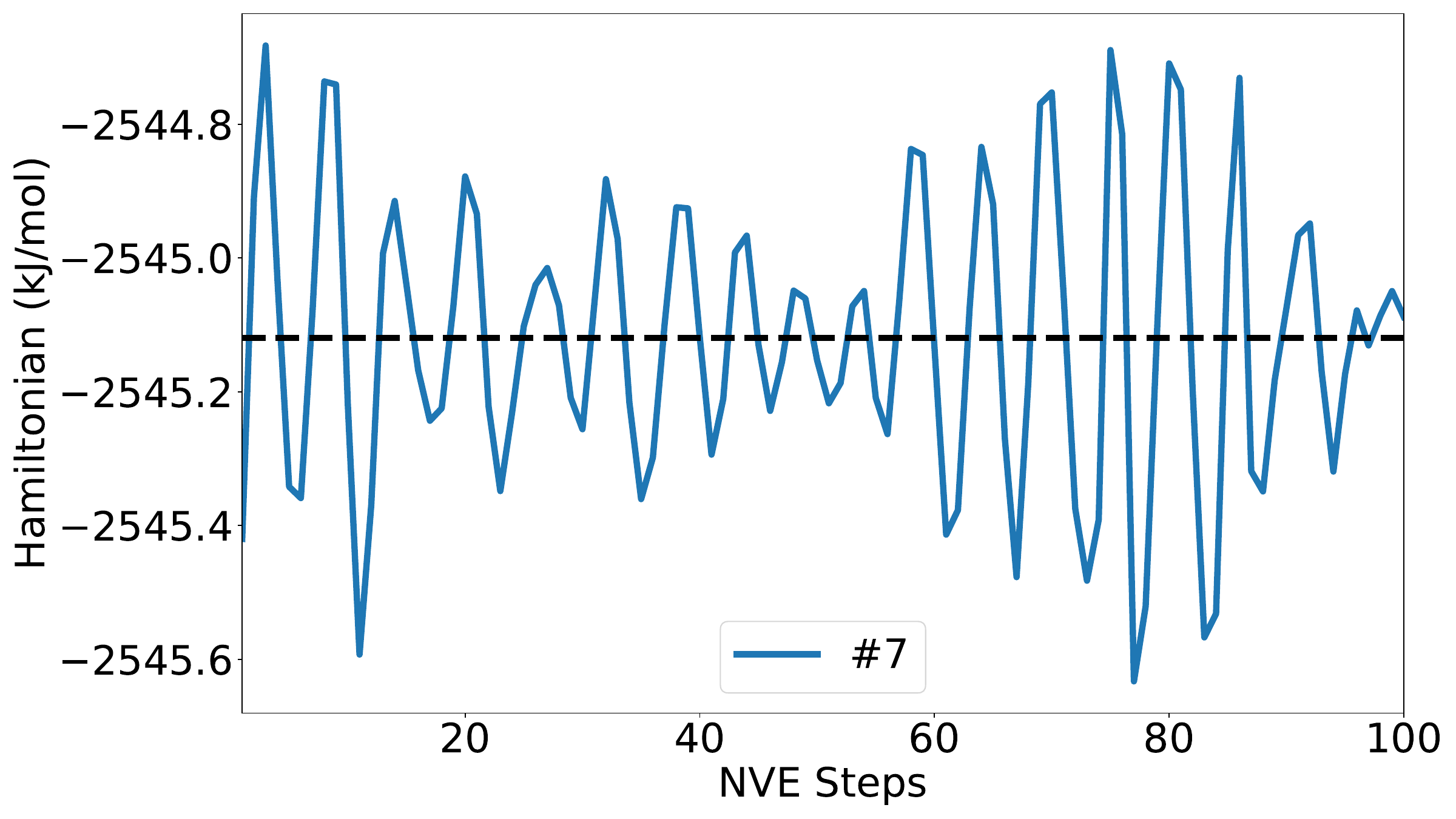}
\end{subfigure}
\begin{subfigure}{0.48\textwidth}
\includegraphics[width=\linewidth]{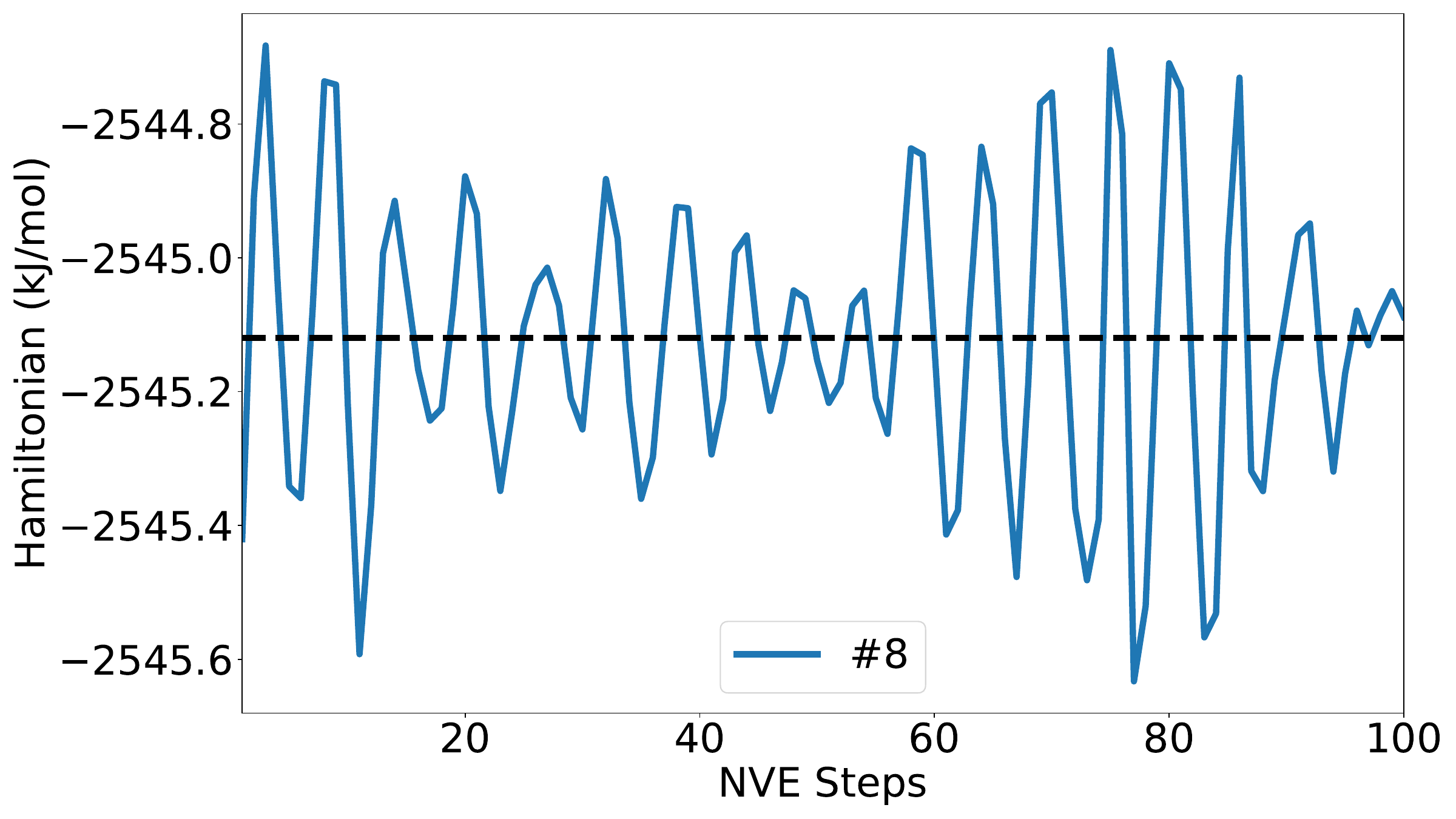}
\end{subfigure}
\caption{Evolution of system Hamiltonians
over 100 time-steps for different deployment strategies.
Each subfigure corresponds to a specific strategy
as detailed in Table~\ref{tb:deploy8}.
Dashed lines indicate the mean Hamiltonian value.}\label{fig:drift}
\end{figure*}
 \else\fi

\textbf{Trajectory Convergence}
Figure~\ref{fig:convergence} depicts the differences per time-step
in potential energy ($U$), kinetic energy ($K$), and Hamiltonian ($H$)
between the baseline and other implementations.
The energies maintained convergence throughout the 100 time-steps,
with numerical differences approaching the theoretical limit of fp32 precision
(approximately the 6th or 7th significant figure).
The spatial coordinates, recorded in the PDB format,
exhibited consistency up to three decimal places
with a maximum deviation of $0.001$~\AA{} on all trajectories.

\ifdefined\InlineFloatEnv
\begin{figure*}[htbp]
\begin{subfigure}{0.48\textwidth}
\includegraphics[width=\linewidth]{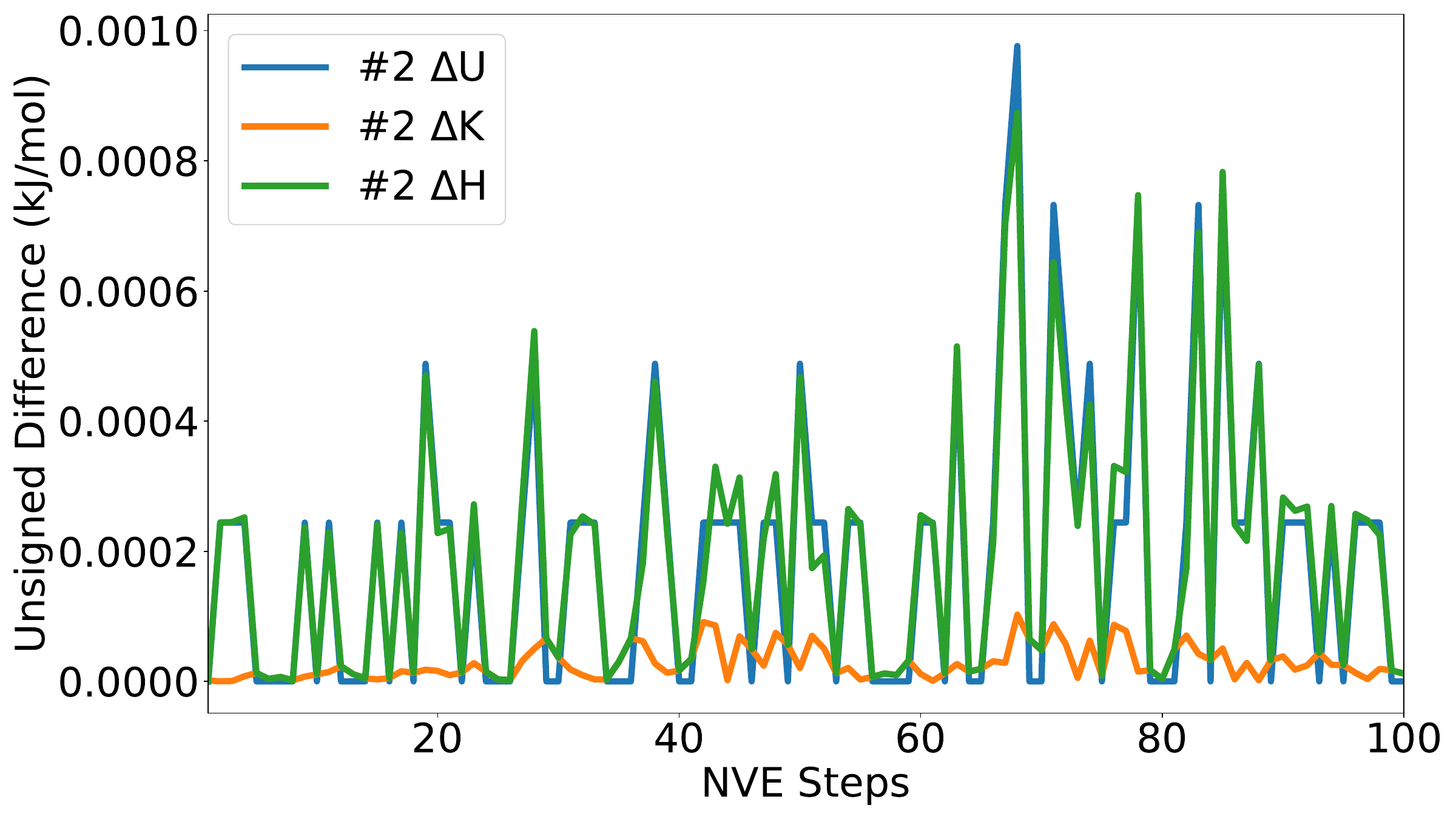}
\end{subfigure}
\\
\begin{subfigure}{0.48\textwidth}
\includegraphics[width=\linewidth]{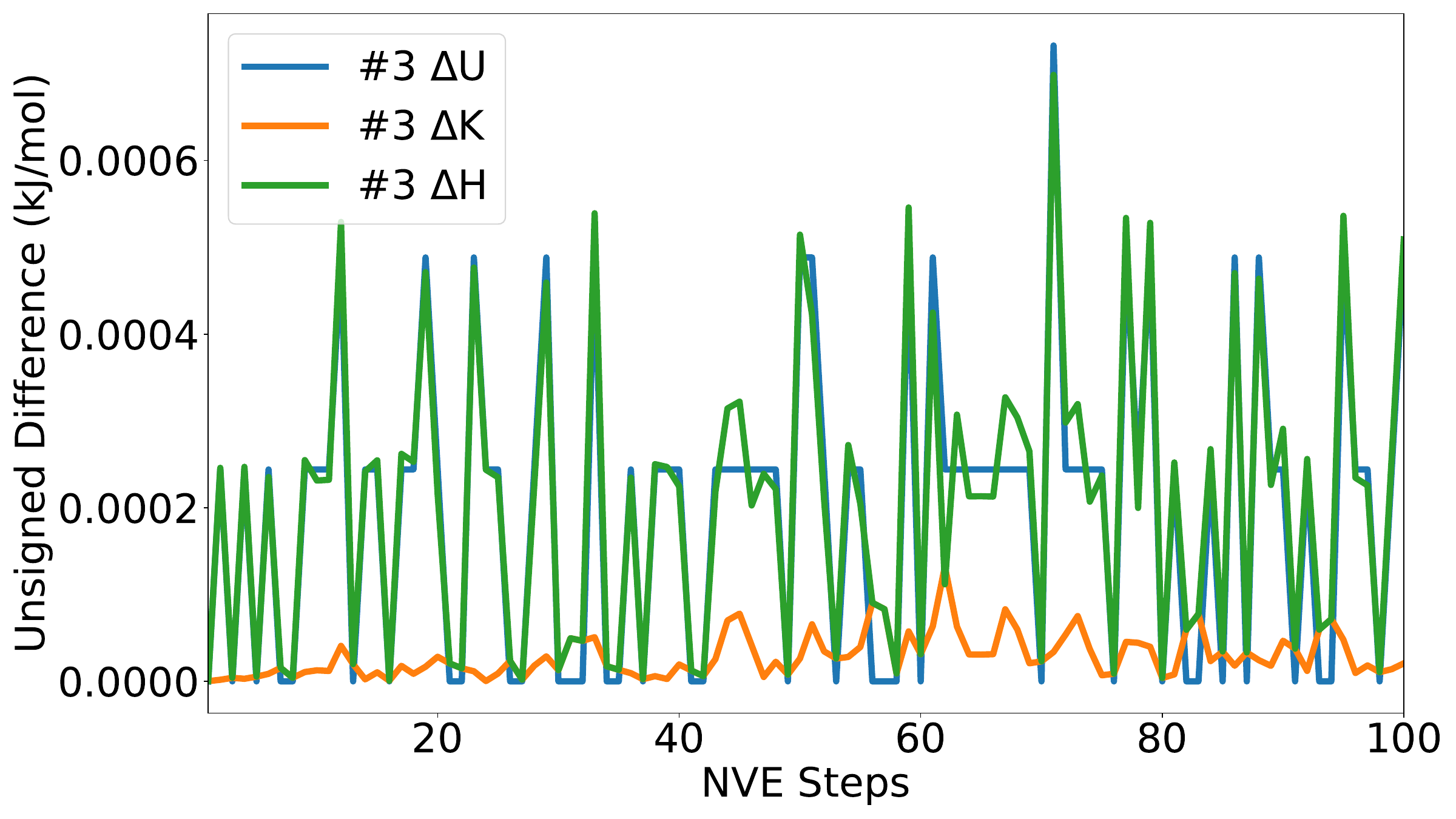}
\end{subfigure}
\begin{subfigure}{0.48\textwidth}
\includegraphics[width=\linewidth]{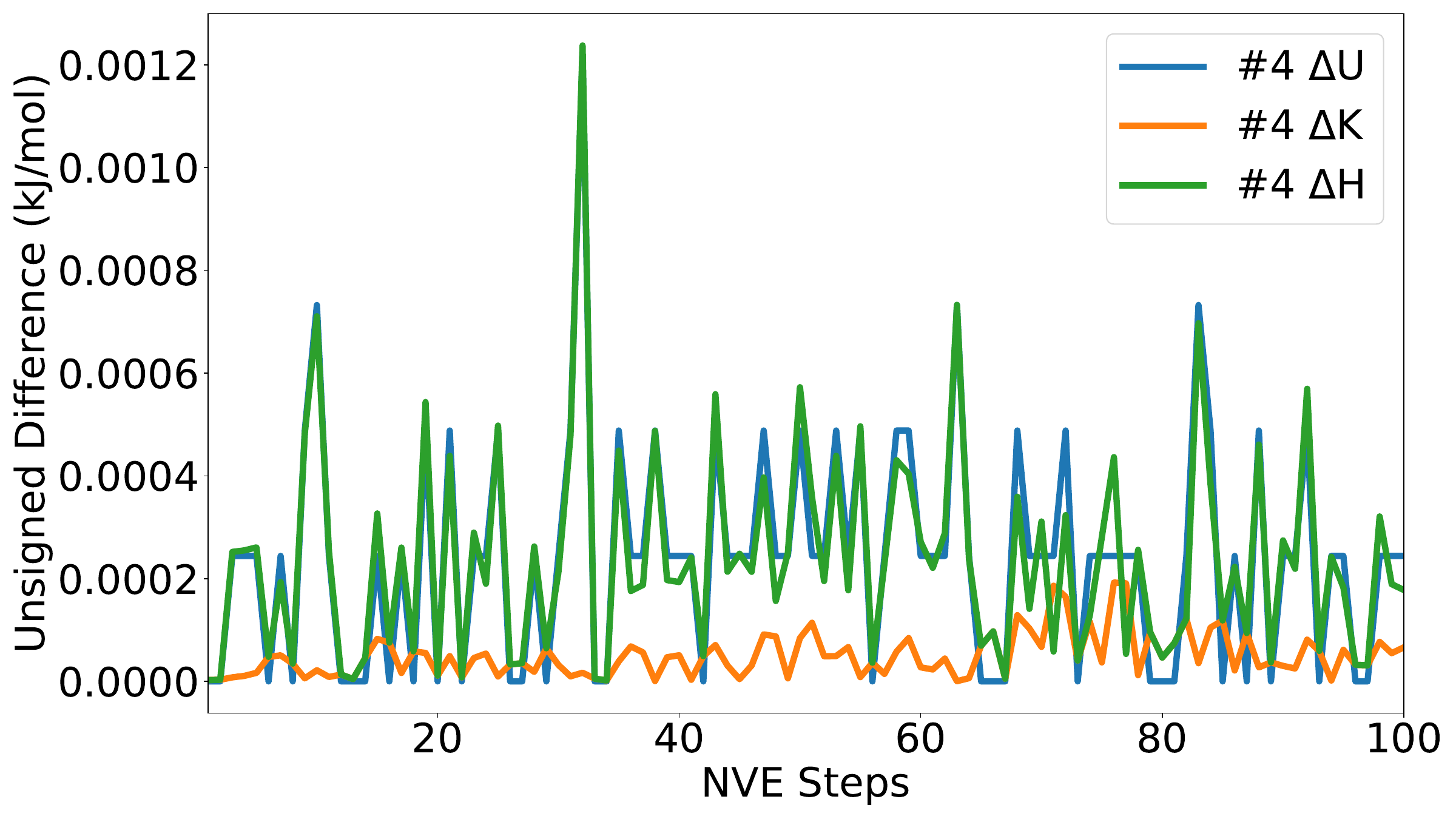}
\end{subfigure}
\\
\begin{subfigure}{0.48\textwidth}
\includegraphics[width=\linewidth]{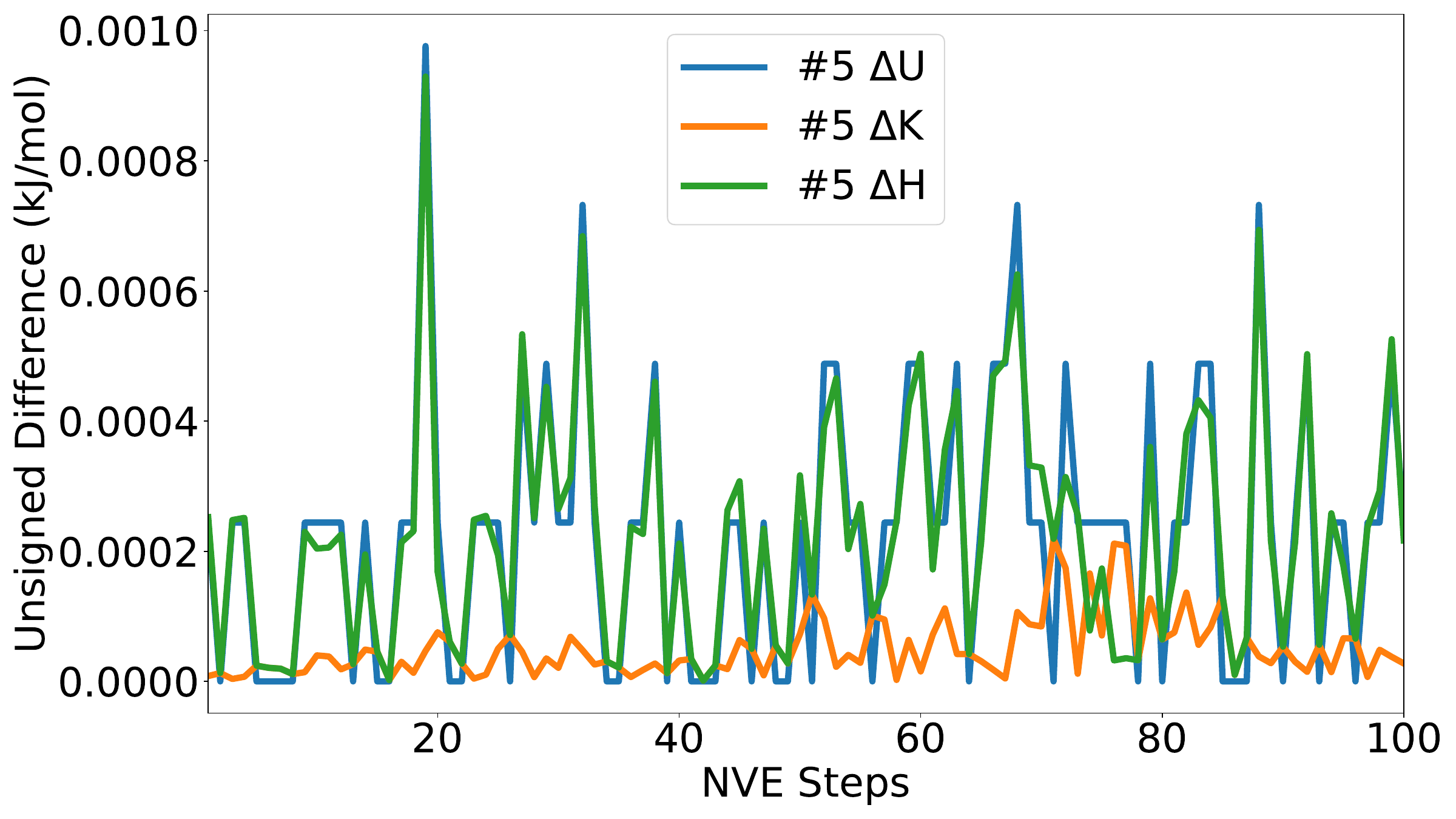}
\end{subfigure}
\begin{subfigure}{0.48\textwidth}
\includegraphics[width=\linewidth]{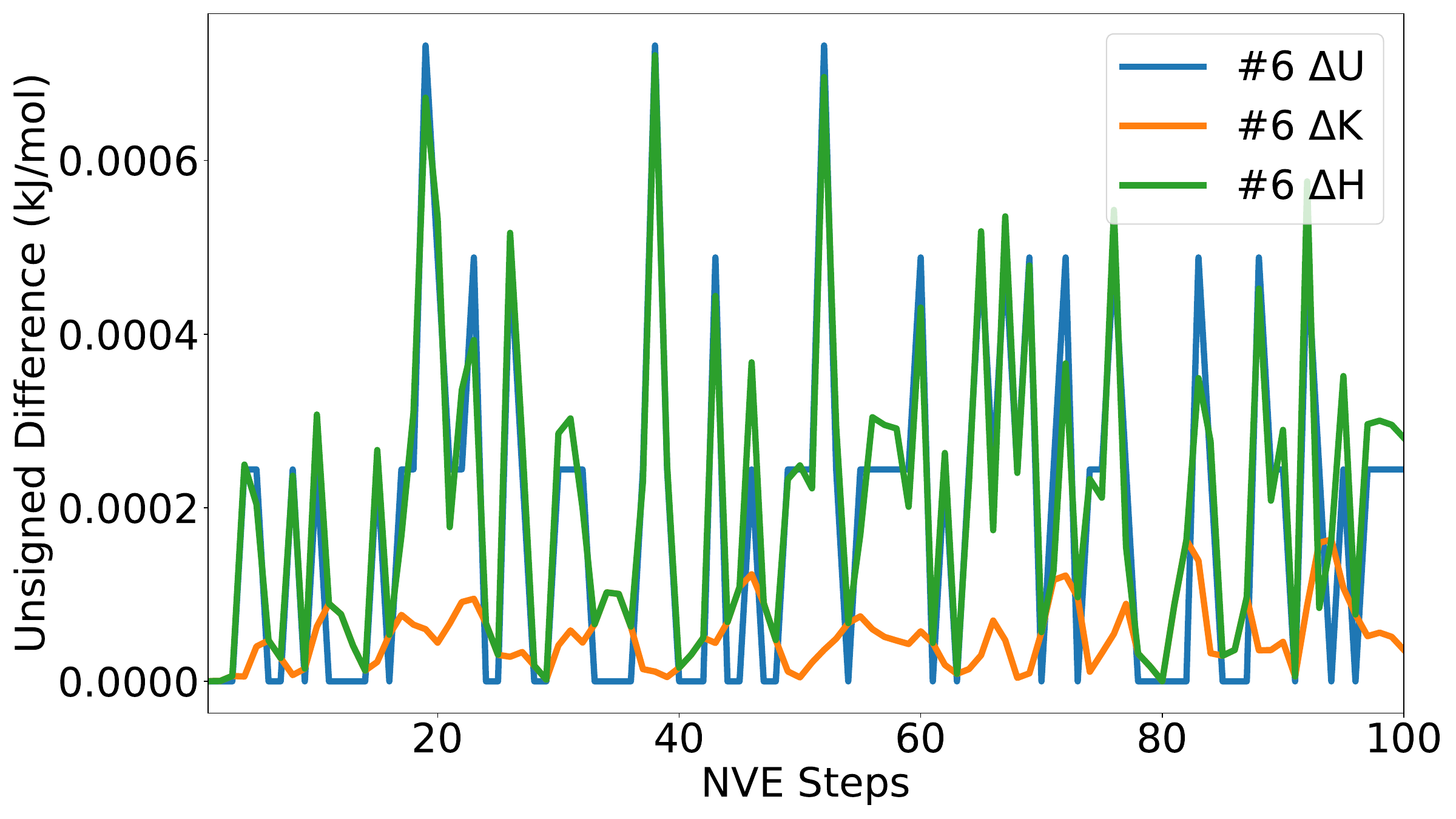}
\end{subfigure}
\\
\begin{subfigure}{0.48\textwidth}
\includegraphics[width=\linewidth]{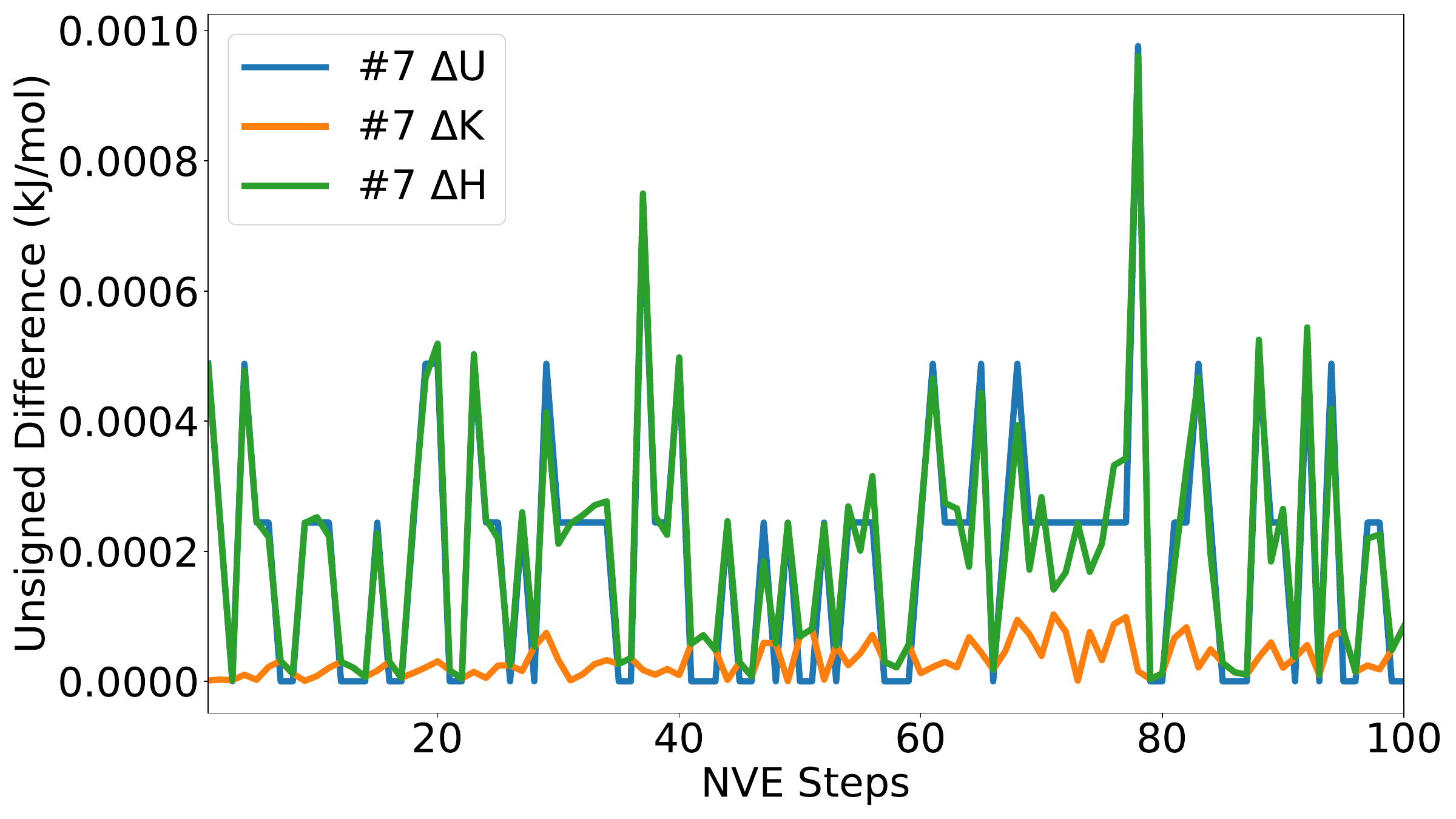}
\end{subfigure}
\begin{subfigure}{0.48\textwidth}
\includegraphics[width=\linewidth]{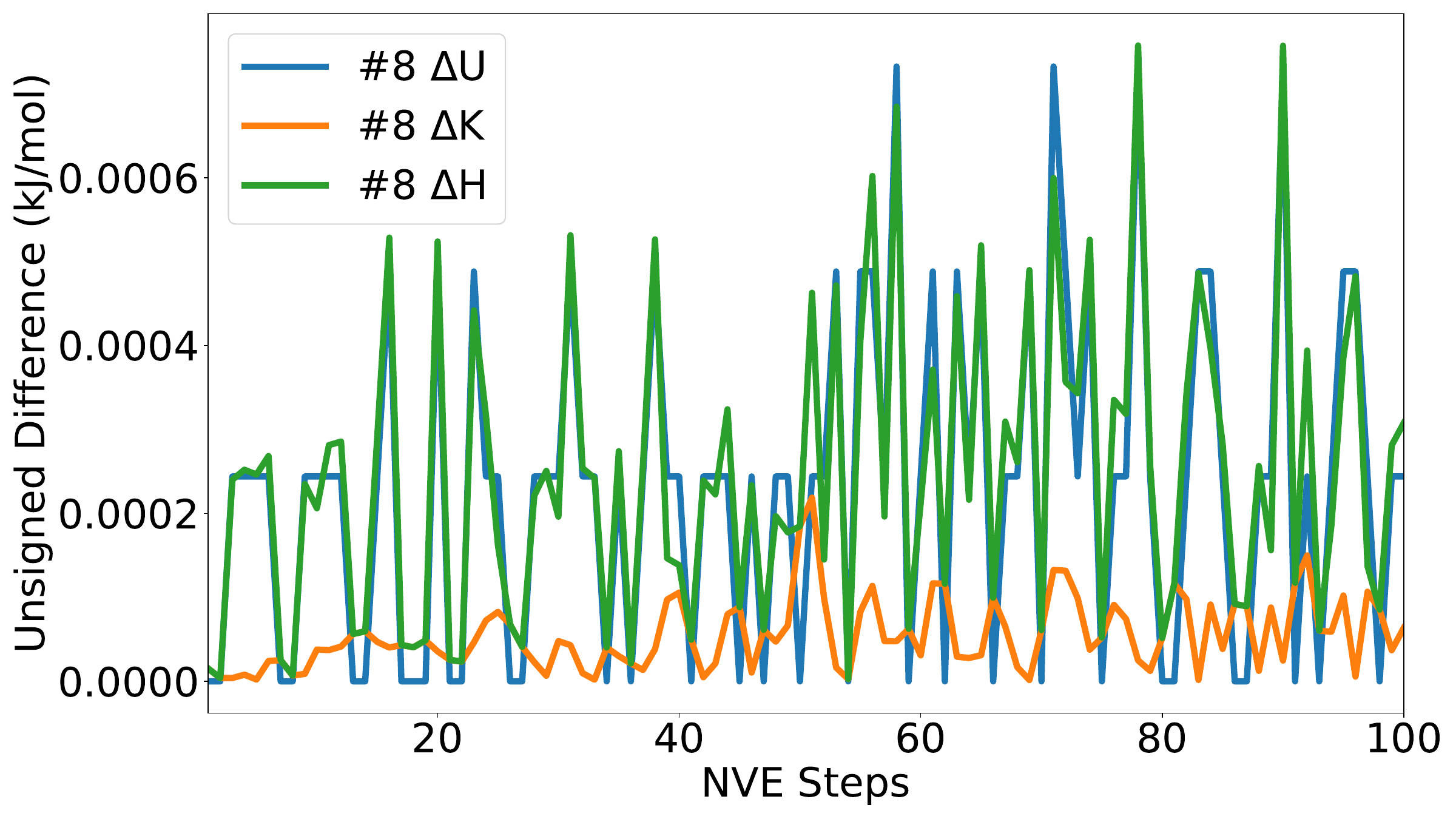}
\end{subfigure}
\caption{Comparison of energies across deployment strategies:
unsigned differences in potential energy ($U$), kinetic energy ($K$), and
Hamiltonian ($H$) relative to the baseline simulation over 100 time-steps.
The number in each subfigure indicates the deployment method
as defined in Table~\ref{tb:deploy8}.}\label{fig:convergence}
\end{figure*}
 \else\fi

\textbf{Errors in Forces}
Using the baseline trajectory, we recalculated the potential energies and forces.
Figure~\ref{fig:reruns} illustrates the unsigned error in potential energies
and root mean square deviation (RMSD) of the 27 force components
relative to the baseline values.
These results corroborate that the differences are minimal,
reaching the inherent precision limit of fp32 arithmetic.

\ifdefined\InlineFloatEnv
\begin{figure*}[htbp]
\begin{subfigure}{0.48\textwidth}
\includegraphics[width=\linewidth]{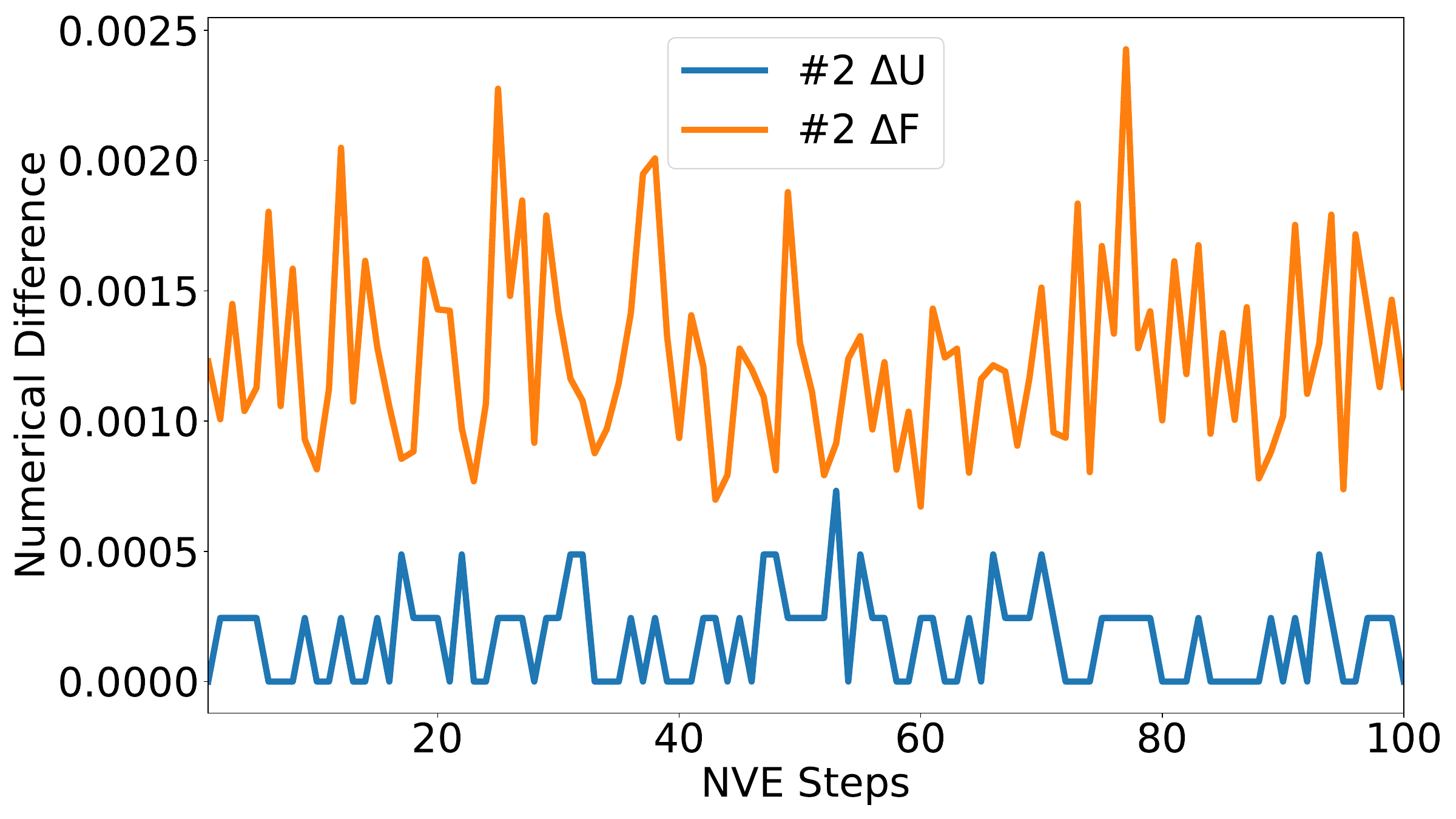}
\end{subfigure}
\\
\begin{subfigure}{0.48\textwidth}
\includegraphics[width=\linewidth]{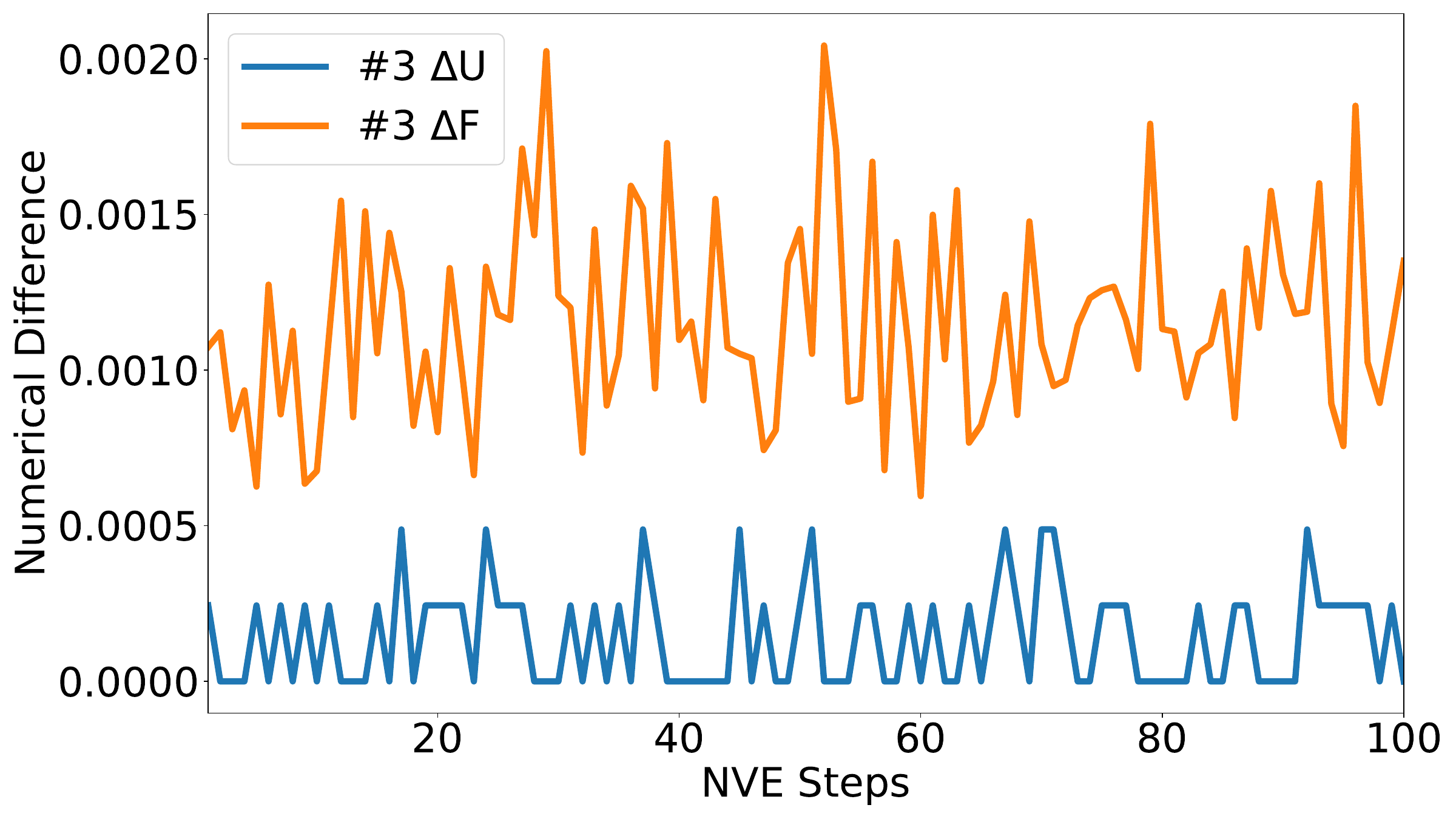}
\end{subfigure}
\begin{subfigure}{0.48\textwidth}
\includegraphics[width=\linewidth]{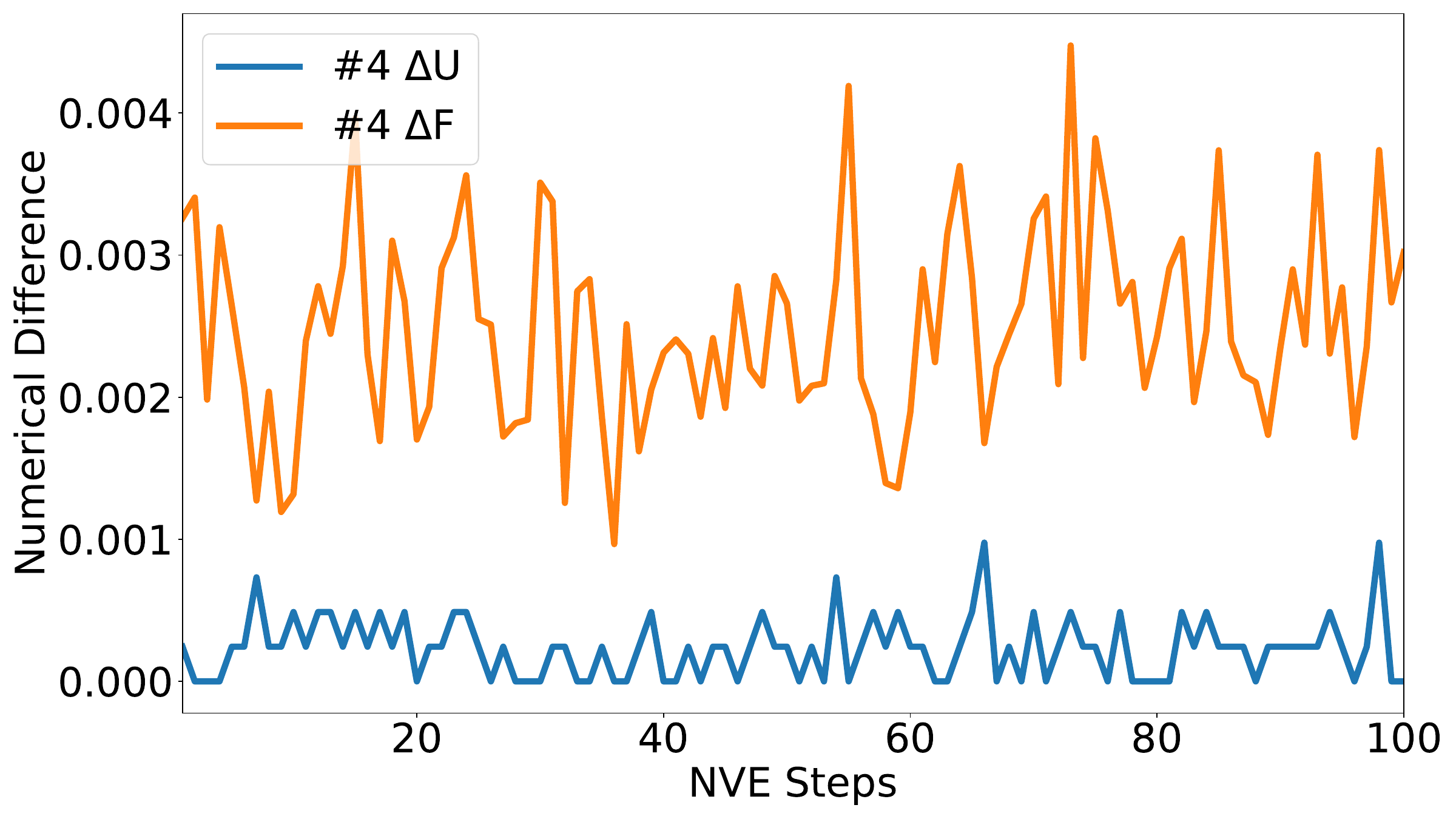}
\end{subfigure}
\\
\begin{subfigure}{0.48\textwidth}
\includegraphics[width=\linewidth]{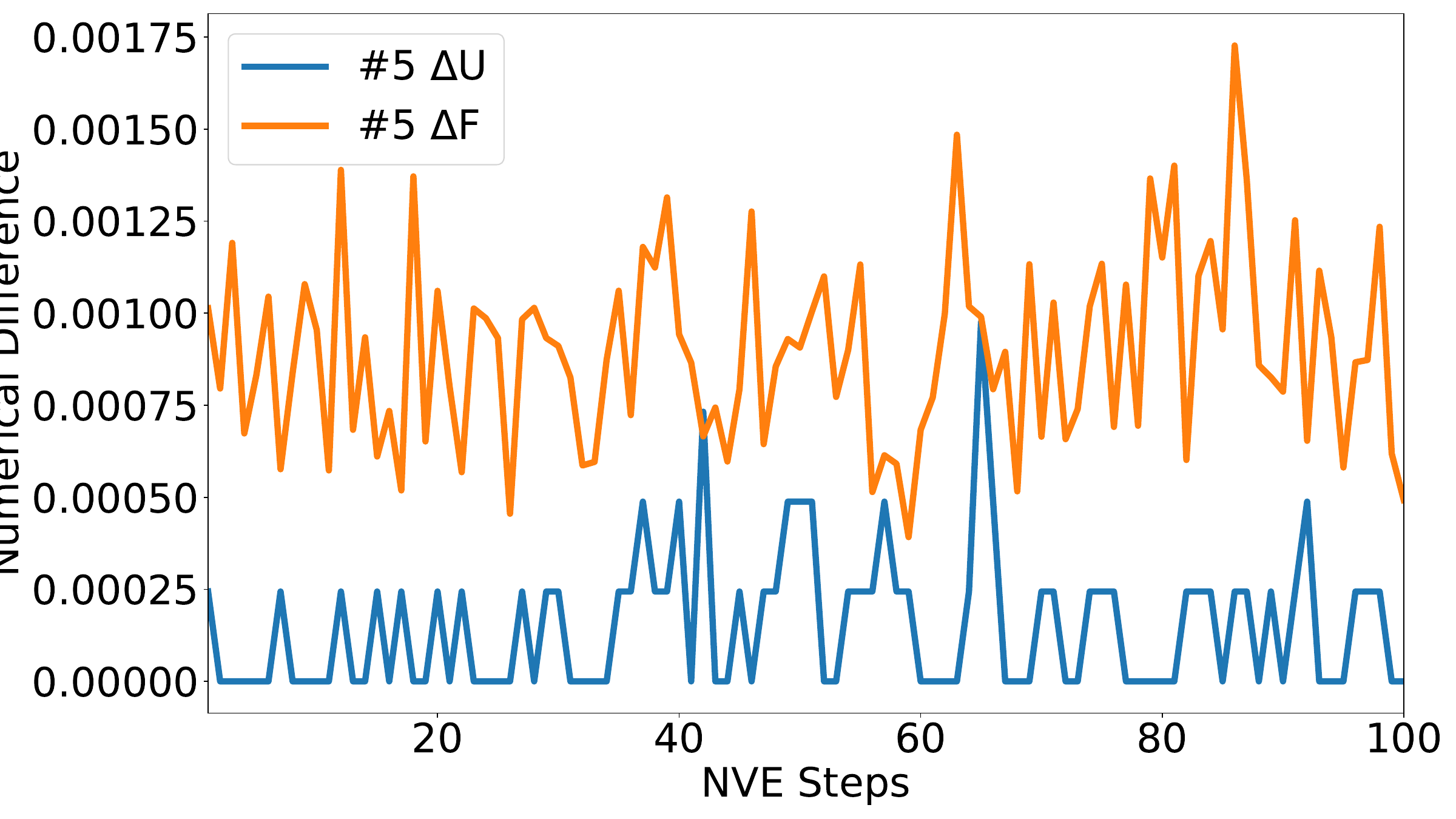}
\end{subfigure}
\begin{subfigure}{0.48\textwidth}
\includegraphics[width=\linewidth]{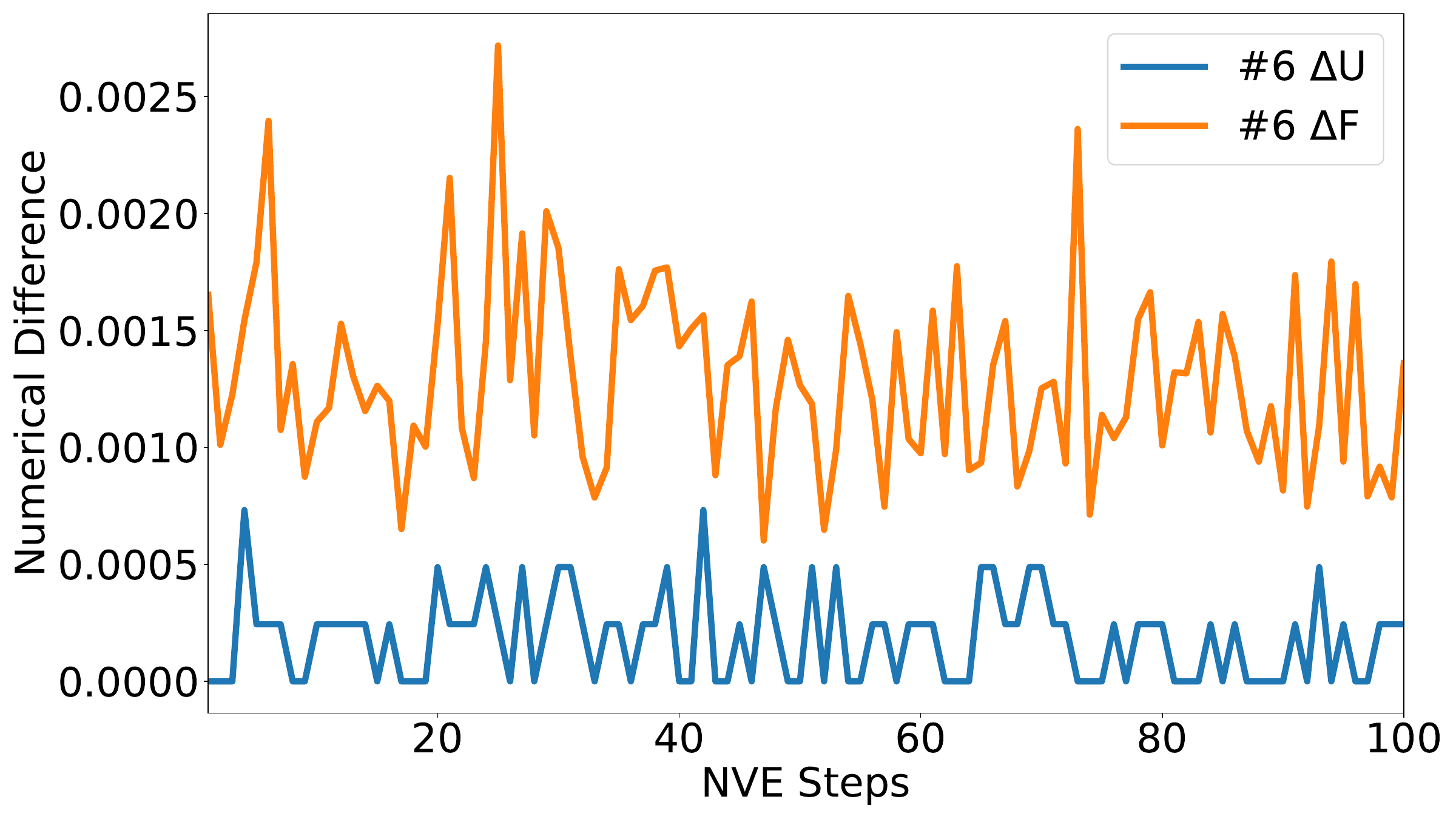}
\end{subfigure}
\\
\begin{subfigure}{0.48\textwidth}
\includegraphics[width=\linewidth]{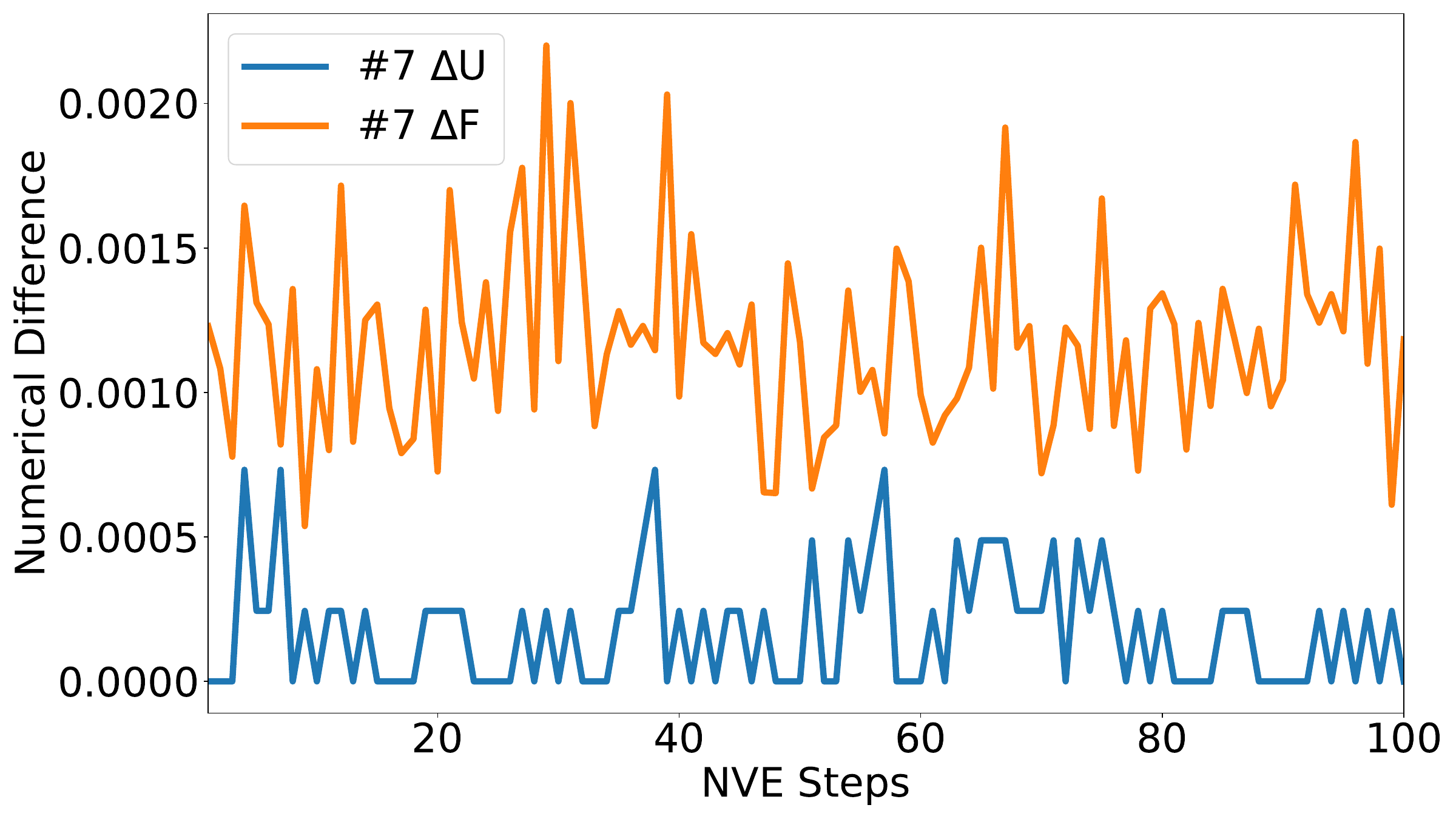}
\end{subfigure}
\begin{subfigure}{0.48\textwidth}
\includegraphics[width=\linewidth]{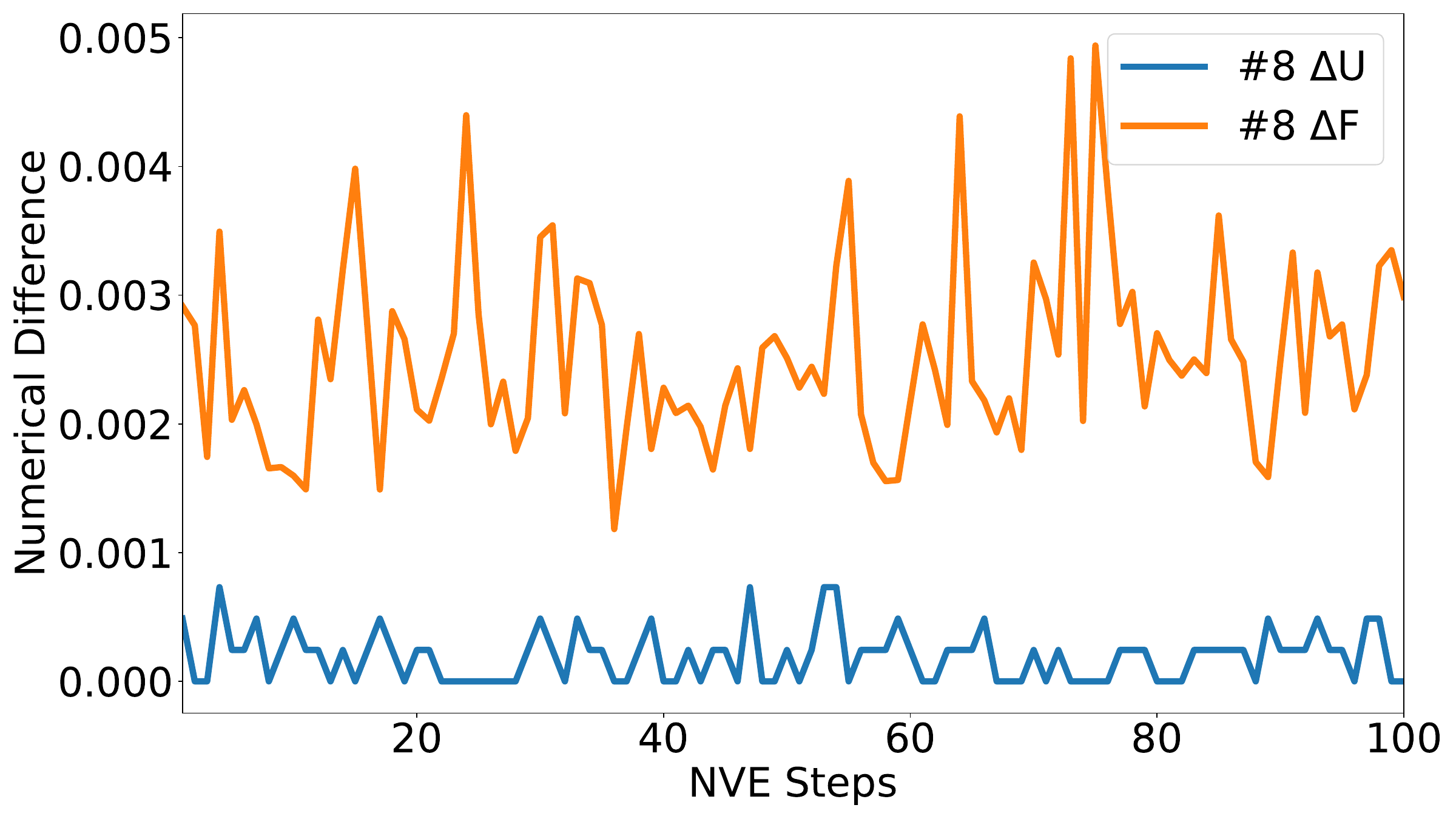}
\end{subfigure}
\caption{The unsigned differences in potential energies
and root mean square deviation of forces (in kJ/mol and nm)
compared to the baseline,
obtained by recalculating an identical PDB trajectory
across deployment methods.
The number in each subfigure denotes the deployment method
as defined in Table~\ref{tb:deploy8}.}
\label{fig:reruns}
\end{figure*}
 \else\fi

The comparative performance of the eight strategies is summarized
in Table~\ref{tb:speed8}.
Performance benchmarks were conducted
using a Langevin integrator with a friction
of $0.1$~ps\textsuperscript{-1} over 1000 time-steps,
generating 10 trajectory frames.
For small-scale test cases such as a single ethanol molecule,
where kernel launch overhead dominates computational cost,
CUDA Graph significantly enhanced performance.
Comparisons between implementations 1 and 3, as well as 5 and 7,
demonstrate that direct inference via C++ APIs achieves superior performance
with reduced launch overhead compared to their Python counterparts.
However, these benefits of reduced overhead are expected to diminish
with increasing system size, a phenomenon previously observed in TorchMD-NET.
Notably, \texttt{\torchcompile} exhibited significant performance advantages,
as evidenced by comparisons between implementations 1 and 4,
and between 5 and 8.

\ifdefined\InlineFloatEnv
\begin{table*}[htbp]
\caption{Performance comparison of different deployment
and compilation strategies in ethanol simulations.
The relative speeds are normalized
to the baseline implementation (\#1).}\label{tb:speed8}
\begin{tabular}{rrrr}
\hline
No. & ms/step & $10^6$ steps/day & Relative Speed \\\hline
1   &    3.97 &             21.8 &            $1$ \\
2   &    7.40 &             11.7 &         $0.54$ \\
3   &    5.58 &             15.5 &         $0.71$ \\
4   &    3.07 &             28.1 &          $1.3$ \\
5   &   0.668 &              129 &          $5.9$ \\
6   &    1.15 &             75.4 &          $3.5$ \\
7   &   0.915 &             94.5 &          $4.3$ \\
8   &   0.486 &              178 &          $8.2$ \\\hline
\end{tabular}
\end{table*}
 \else\fi
 
\subsection{Example: AIMD Simulation}

The versatility of the callback mechanism extends
beyond PyTorch tensors to accommodate diverse data containers.
To demonstrate this flexibility,
we implemented an AIMD simulation using PySCF/GPU4PySCF \cite{Sun2020,Wu2024,Li2024,Lehtola2018}
with the B3LYP functional and D3BJ dispersion correction.
The simulation utilizes \emph{NumPyForce}, an OpenMM Force plugin we developed
alongside \emph{TorchForce} to establish a bridge
between OpenMM and quantum chemistry software packages.
All relevant implementation files are available on GitHub.

In this quantum mechanical context,
where automatic differentiation via backpropagation is not available,
forces or gradients must be explicitly provided to the MD engine.
While NumPy lacks native CUDA support,
the overhead from device-to-device data transfer
and floating-point conversion
(between fp32 and fp64) proves negligible in AIMD simulations,
where quantum chemical force calculations typically
dominate the computational cost.
 
\subsection{Extensibility to Other MD Engines}

The proposed callback mechanism demonstrates broad compatibility
with other molecular dynamics engines,
such as Tinker \cite{Rackers2018} and LAMMPS \cite{Thompson2022},
even when these packages do not natively initialize a Python interpreter.
Based on our implementation experience,
incorporating a ``callback Python energy term''
would require comparable code modifications
in their respective source files.
Furthermore, the initialization of the Python interpreter would necessitate
only minimal additional changes to other components of the codebase,
making it a straightforward extension.

Taking Tinker as an example,
initializing a Python interpreter conceptually resembles
the initialization of a Fortran runtime library,
as currently implemented in Tinker9 \cite{Software-Tinker9}.
During program initialization, Tinker9 calls compiler-specific functions:
\verb|_gfortran_set_args| for GFortran-compiled executables,
or \verb|for_rtl_init_| and \verb|for_rtl_finish_|
for initialization and cleanup with the Intel compiler, respectively.
Similarly, implementing Python support would primarily involve
incorporating CPython C-API functions
such as \verb|Py_Initialize| and \verb|Py_Finalize|.
For detailed implementation guidance,
we refer readers to the official CPython documentation \cite{Software-Python-C-API}
and pybind11's \emph{embedding the interpreter} documentation \cite{Software-pybind11-embed}.
  
\section{Conclusion}

In this work, we have presented OpenMM-Python-Force,
a callback mechanism that seamlessly bridges molecular dynamics simulations
with machine learning model inference.
Our evaluation demonstrates that this approach is not only robust and
computationally efficient but also remarkably versatile in its applications.
The applications of this callback mechanism extend well
beyond its initial implementation with PyTorch and OpenMM,
encompassing both classical and ab initio molecular dynamics simulations.
We anticipate that this work will substantially reduce the technical barriers
for integrating various computational backends with MD simulations,
thereby accelerating progress in relevant fields of research.
 
\section*{Acknowledgements}

We extend our gratitude
to Leyuan Wang and Siyuan Liu
for their assistance in identifying the source
of numerical discrepancies between results produced
by \texttt{\torchcompile} and other deployment strategies;
Chi Xu for conducting the code review;
and Sheng Gong, Zhenliang Mu, Zhichen Pu, and Xu Han
for their support in setting up the BAMBOO model.
We also thank Xiaojie Wu
for his contributions to the code interfacing GPU4PySCF with this work.
 
\bibliography{refs}

\ifdefined\InlineFloatEnv\else
\newpage

\begin{table*}[htbp]
\caption{Deployment and compilation strategies evaluated
in ethanol simulations.}\label{tb:deploy8}
\begin{tabular}{c|cr|cr}
\hline
       & No. & Description                   & No. & Description   \\\hline
C++    & 1   & OpenMM Torch (Baseline)       & 5  & 1 + CUDA Graph \\
Python & 2   & native \verb|torch.nn.Module| & 6  & 2 + CUDA Graph \\
Python & 3   & \texttt{\torchjit}            & 7  & 3 + CUDA Graph \\
Python & 4   & \texttt{\torchcompile}        & 8  & 4 + CUDA Graph \\\hline
\end{tabular}
\end{table*}
 
\begin{table*}[htbp]
\caption{Performance comparison of different deployment
and compilation strategies in ethanol simulations.
The relative speeds are normalized
to the baseline implementation (\#1).}\label{tb:speed8}
\begin{tabular}{rrrr}
\hline
No. & ms/step & $10^6$ steps/day & Relative Speed \\\hline
1   &    3.97 &             21.8 &            $1$ \\
2   &    7.40 &             11.7 &         $0.54$ \\
3   &    5.58 &             15.5 &         $0.71$ \\
4   &    3.07 &             28.1 &          $1.3$ \\
5   &   0.668 &              129 &          $5.9$ \\
6   &    1.15 &             75.4 &          $3.5$ \\
7   &   0.915 &             94.5 &          $4.3$ \\
8   &   0.486 &              178 &          $8.2$ \\\hline
\end{tabular}
\end{table*}
 
\begin{figure*}[htbp]
\includegraphics[width=\textwidth]{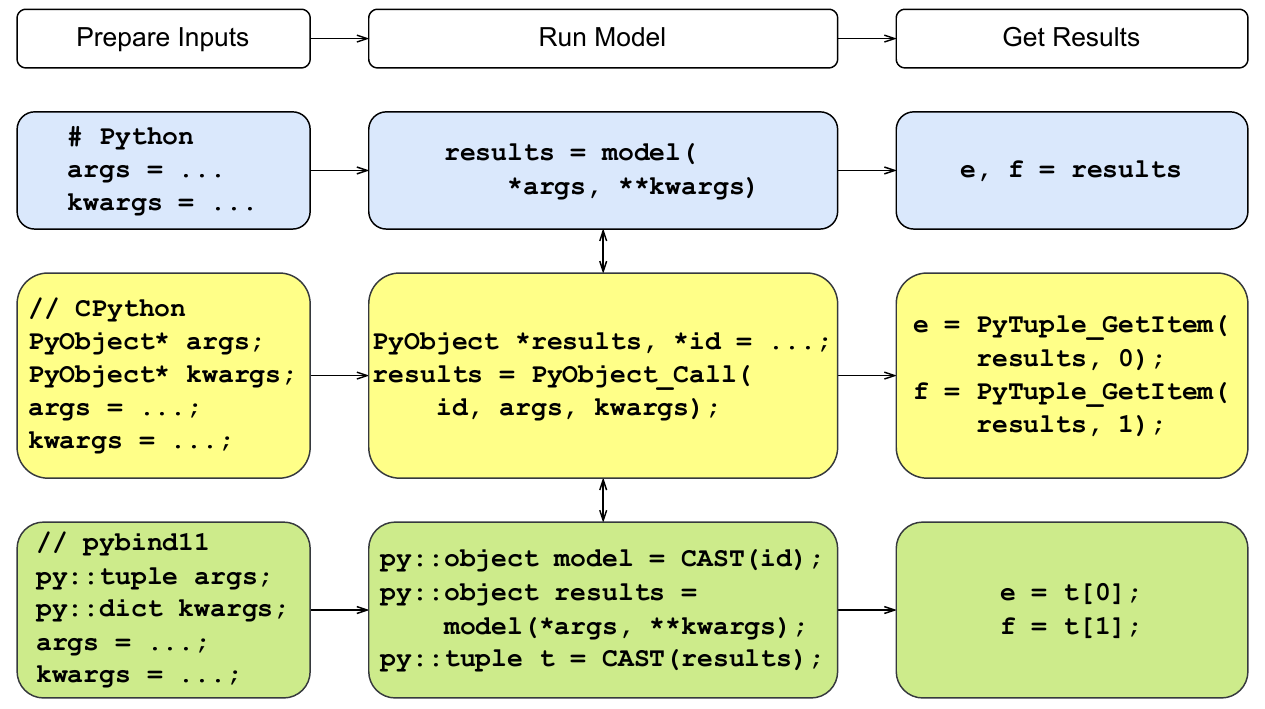}
\caption{Illustration of the Python callback mechanism,
demonstrating the translation between a Python function call
and its corresponding pseudo C/C++ implementations
using either the CPython API or pybind11
(with the C++ namespace \texttt{pybind11} abbreviated
as \texttt{py}).}\label{fig:Callback-C-API-vs-pybind11}
\end{figure*}
 
\begin{figure*}[htbp]
\begin{subfigure}{0.48\textwidth}
\includegraphics[width=\linewidth]{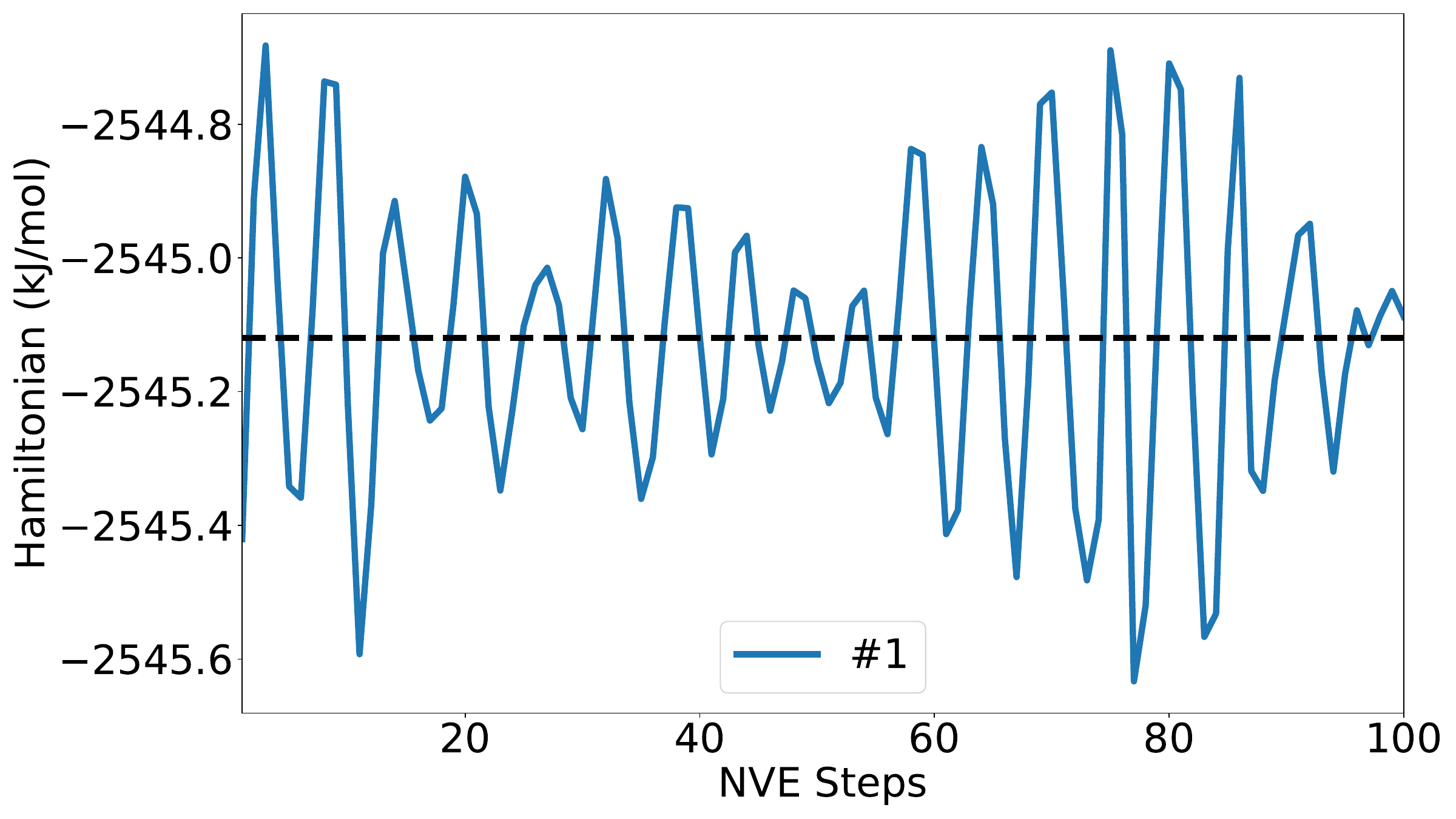}
\end{subfigure}
\begin{subfigure}{0.48\textwidth}
\includegraphics[width=\linewidth]{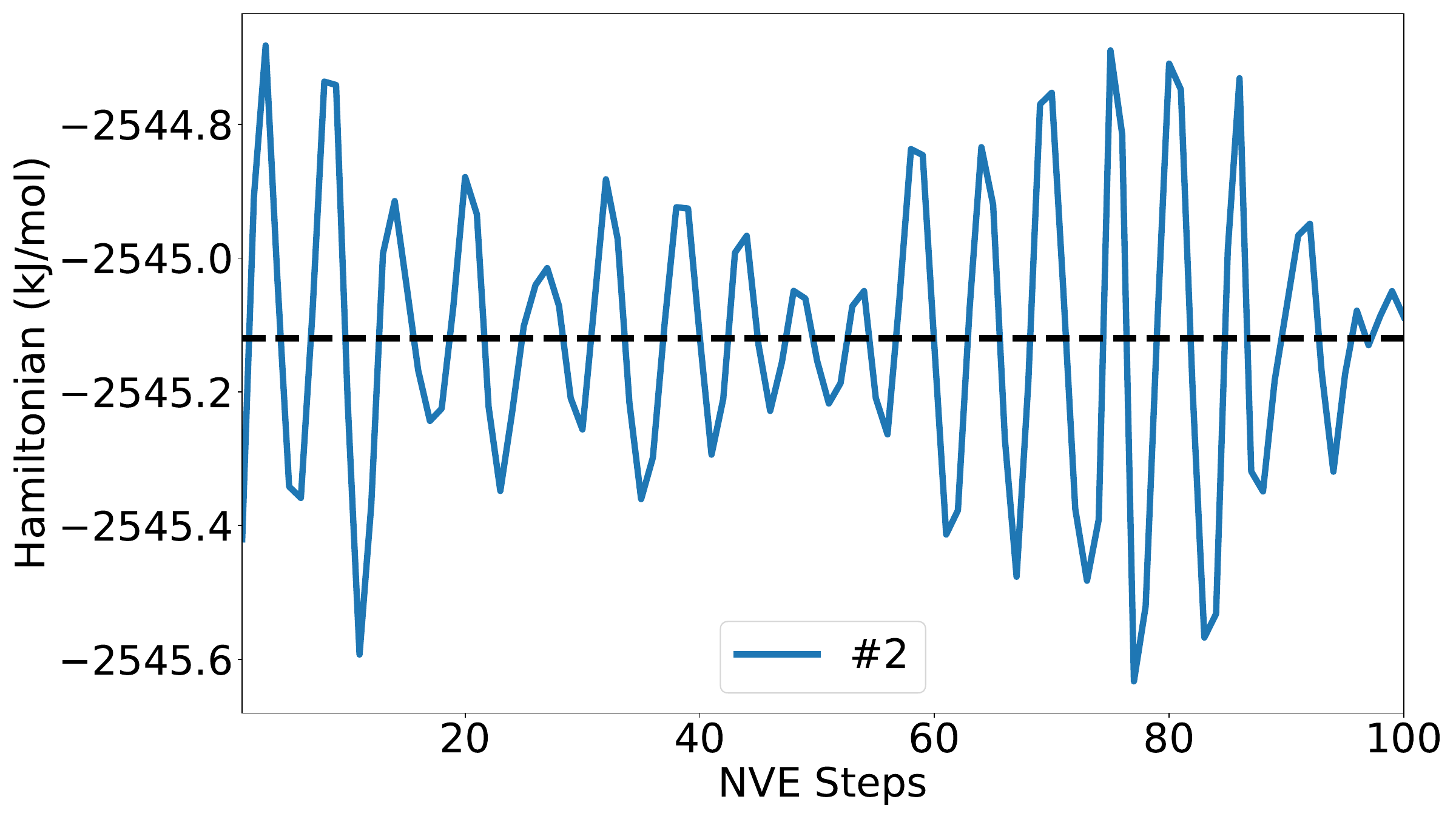}
\end{subfigure}
\\
\begin{subfigure}{0.48\textwidth}
\includegraphics[width=\linewidth]{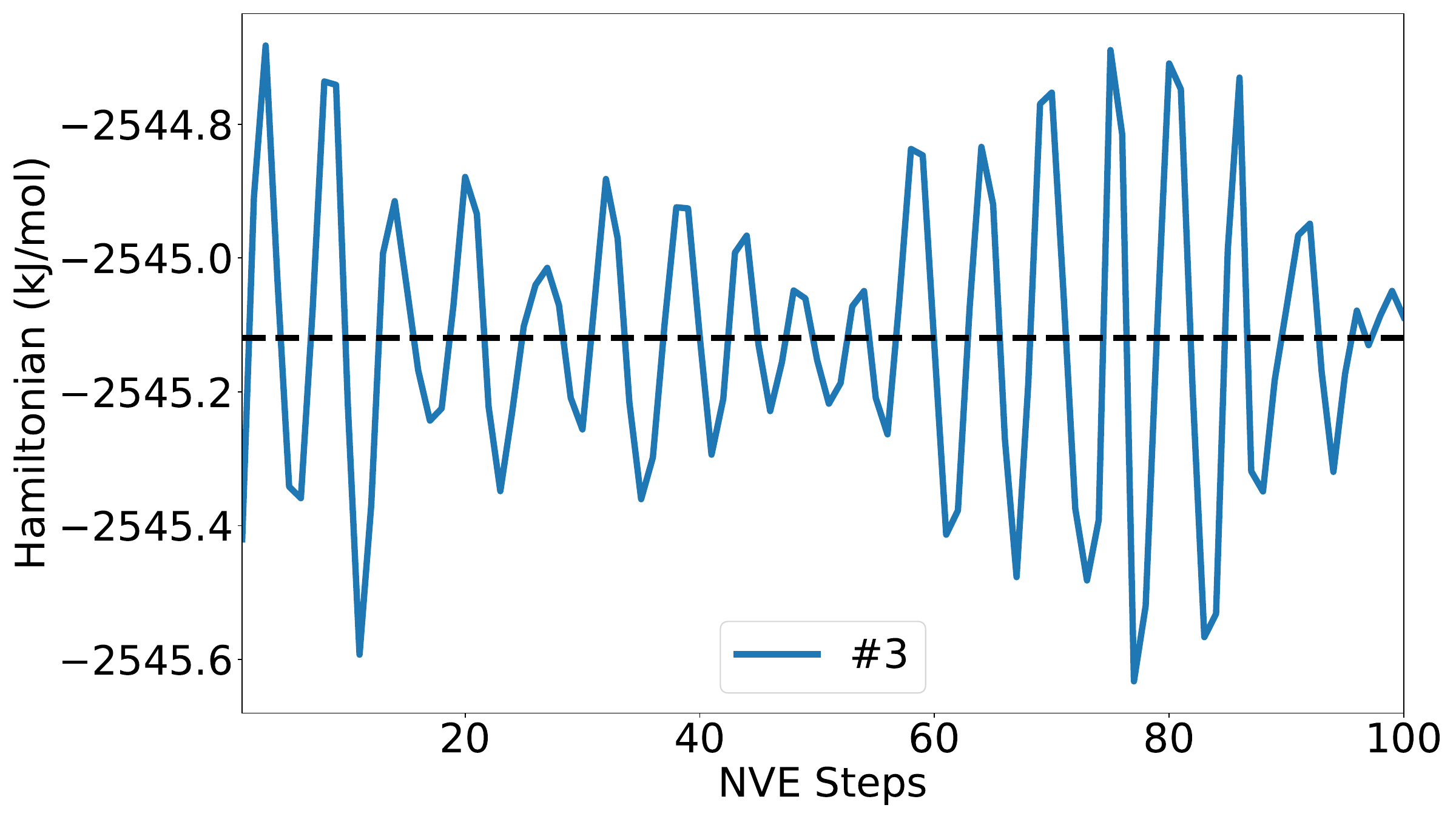}
\end{subfigure}
\begin{subfigure}{0.48\textwidth}
\includegraphics[width=\linewidth]{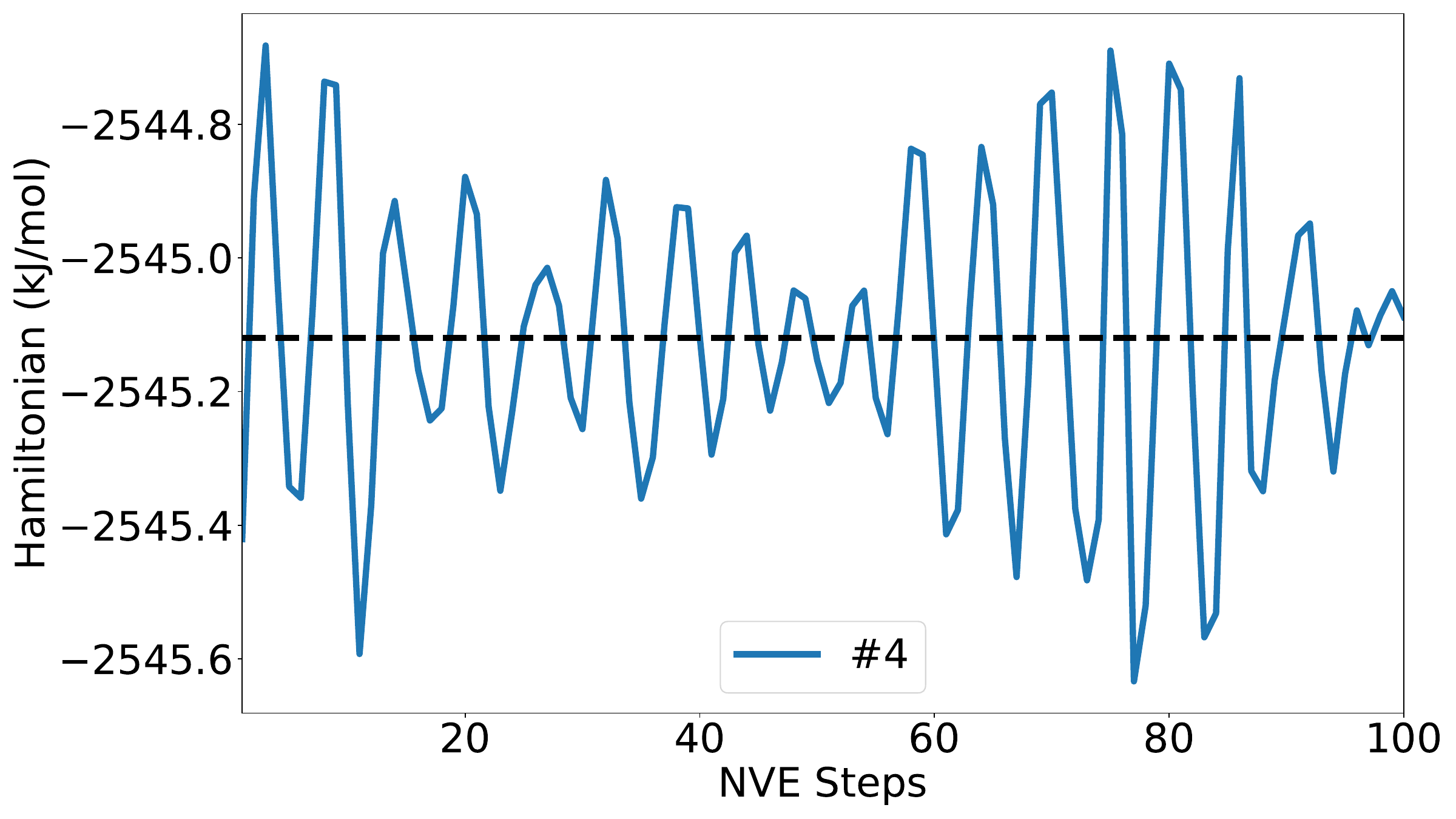}
\end{subfigure}
\\
\begin{subfigure}{0.48\textwidth}
\includegraphics[width=\linewidth]{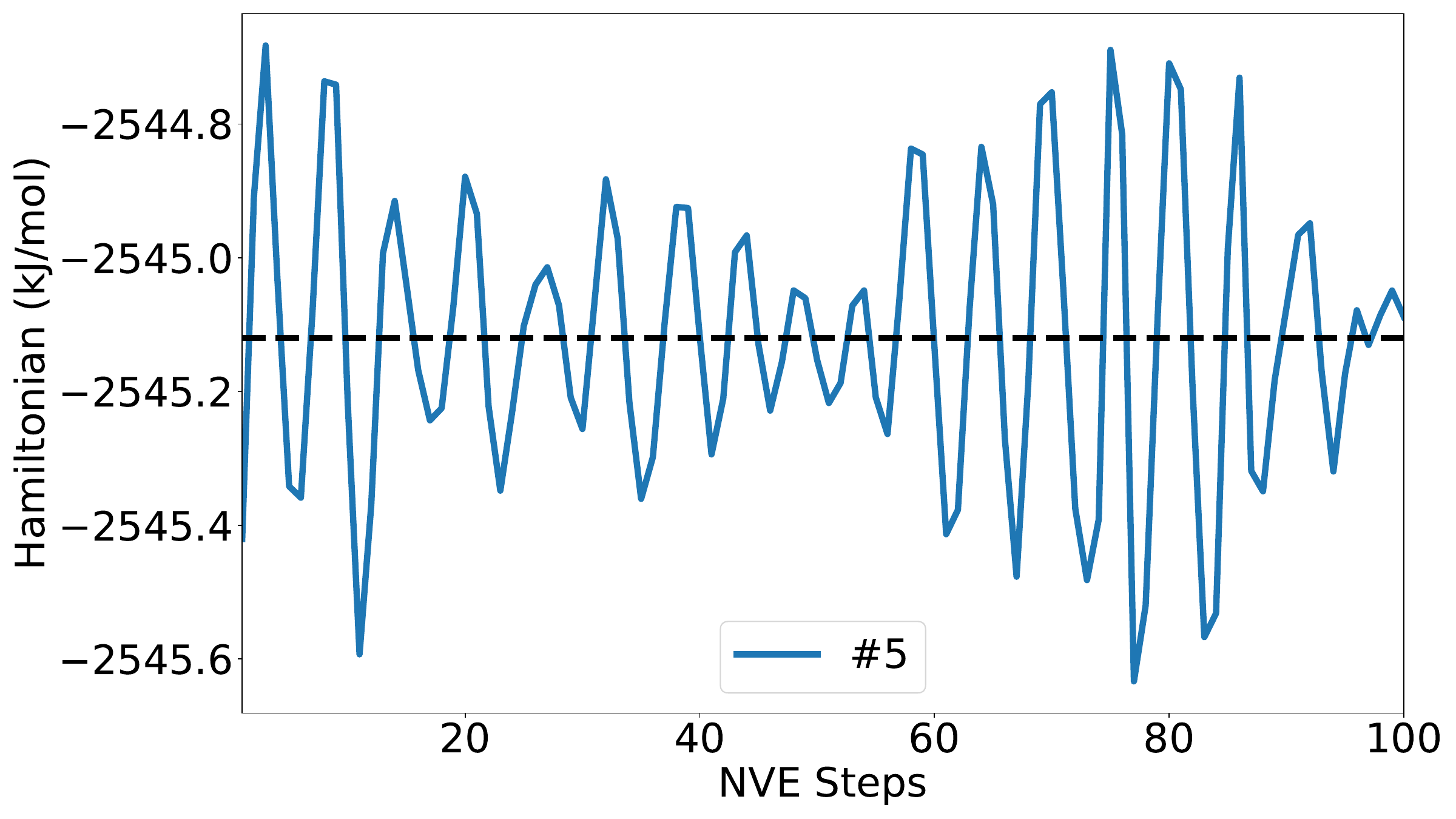}
\end{subfigure}
\begin{subfigure}{0.48\textwidth}
\includegraphics[width=\linewidth]{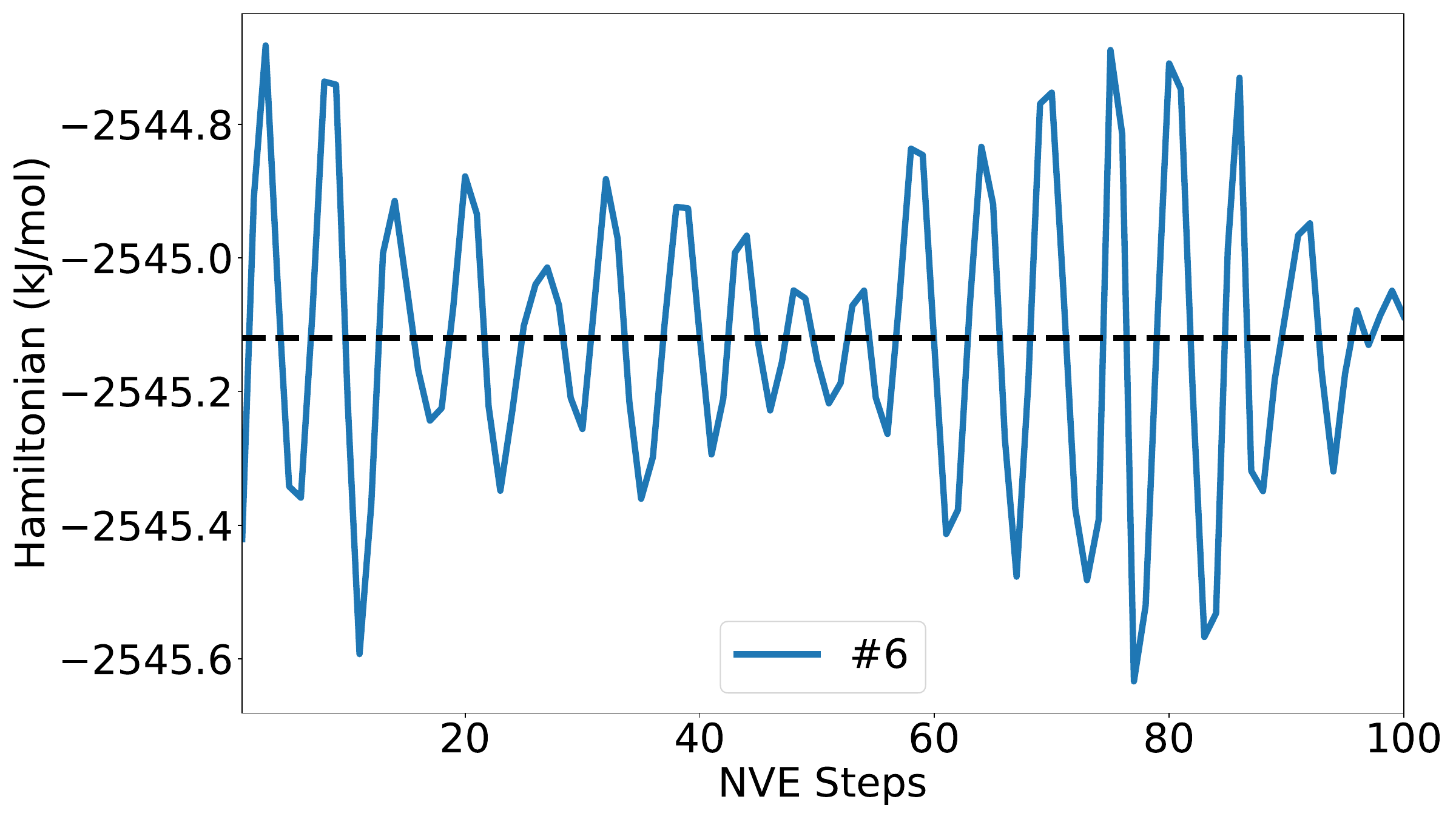}
\end{subfigure}
\\
\begin{subfigure}{0.48\textwidth}
\includegraphics[width=\linewidth]{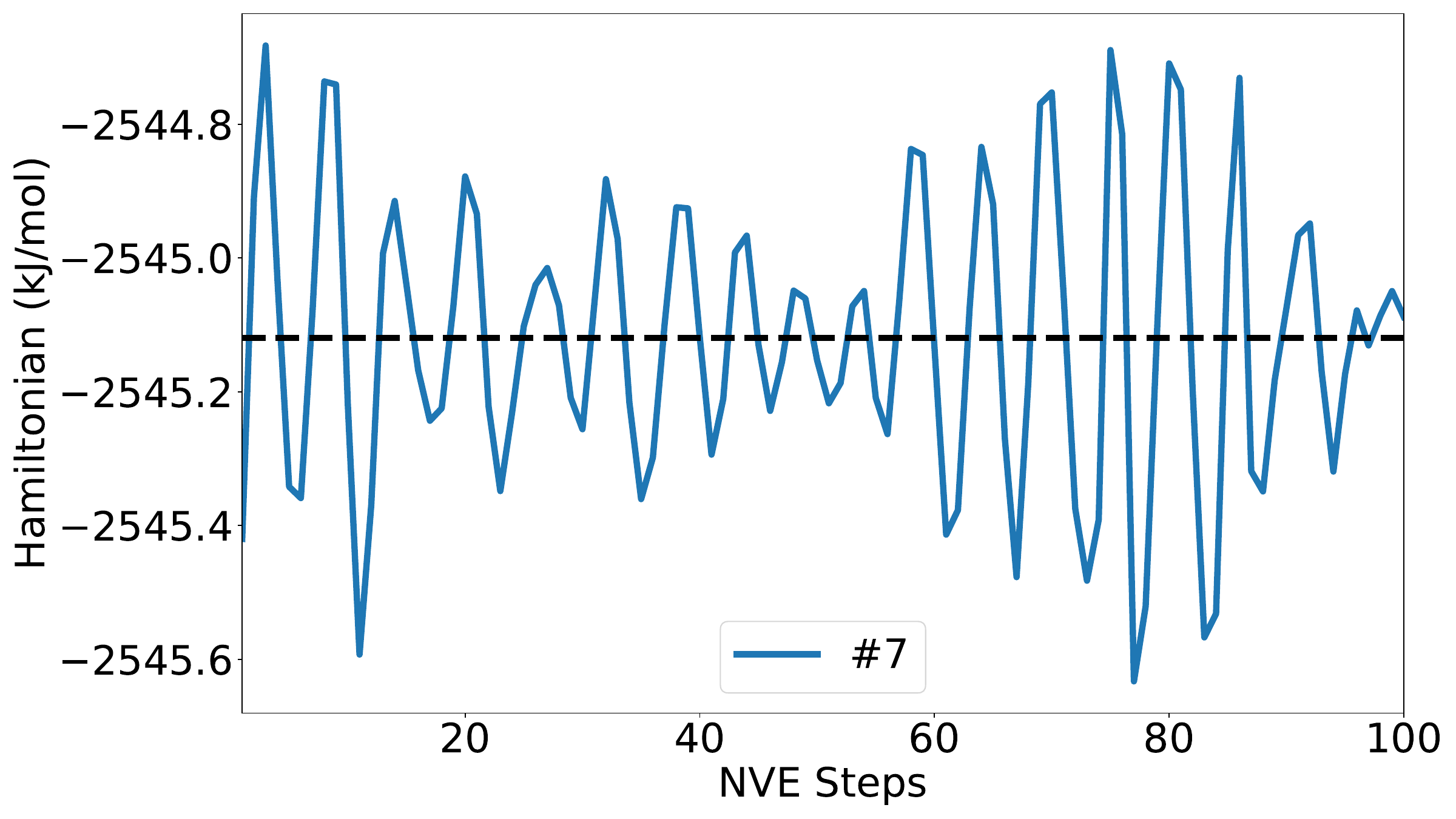}
\end{subfigure}
\begin{subfigure}{0.48\textwidth}
\includegraphics[width=\linewidth]{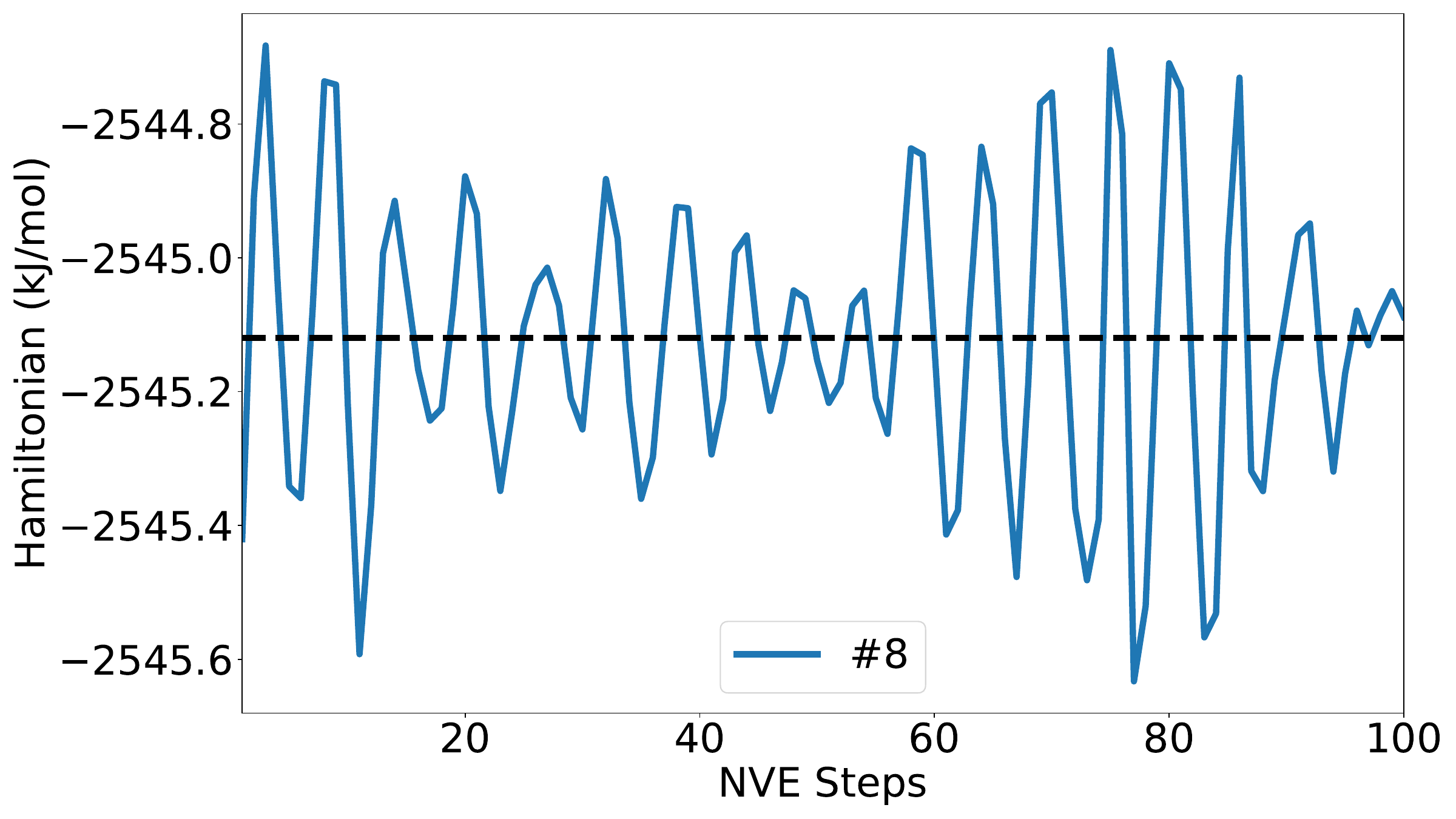}
\end{subfigure}
\caption{Evolution of system Hamiltonians
over 100 time-steps for different deployment strategies.
Each subfigure corresponds to a specific strategy
as detailed in Table~\ref{tb:deploy8}.
Dashed lines indicate the mean Hamiltonian value.}\label{fig:drift}
\end{figure*}
 
\begin{figure*}[htbp]
\begin{subfigure}{0.48\textwidth}
\includegraphics[width=\linewidth]{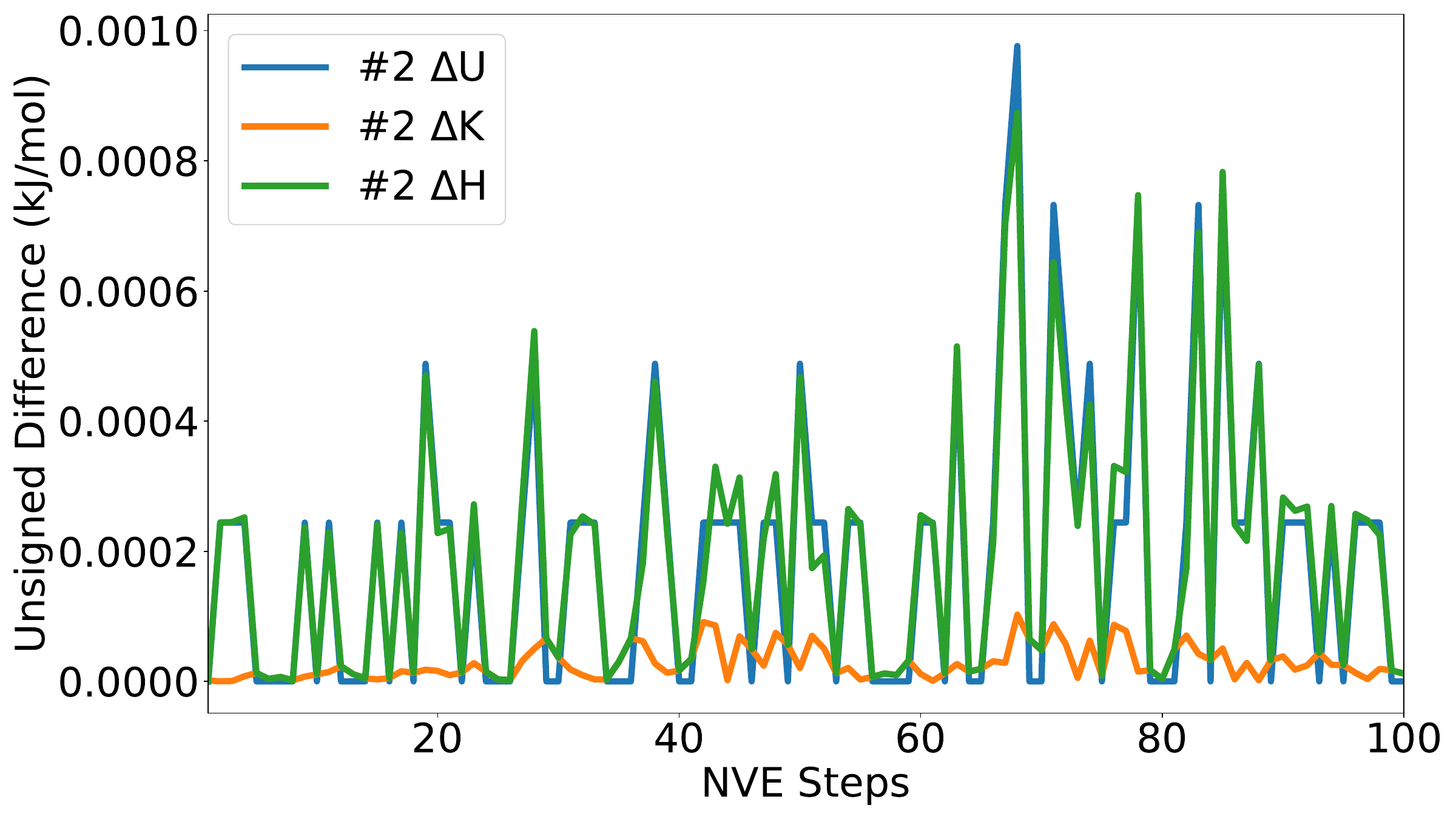}
\end{subfigure}
\\
\begin{subfigure}{0.48\textwidth}
\includegraphics[width=\linewidth]{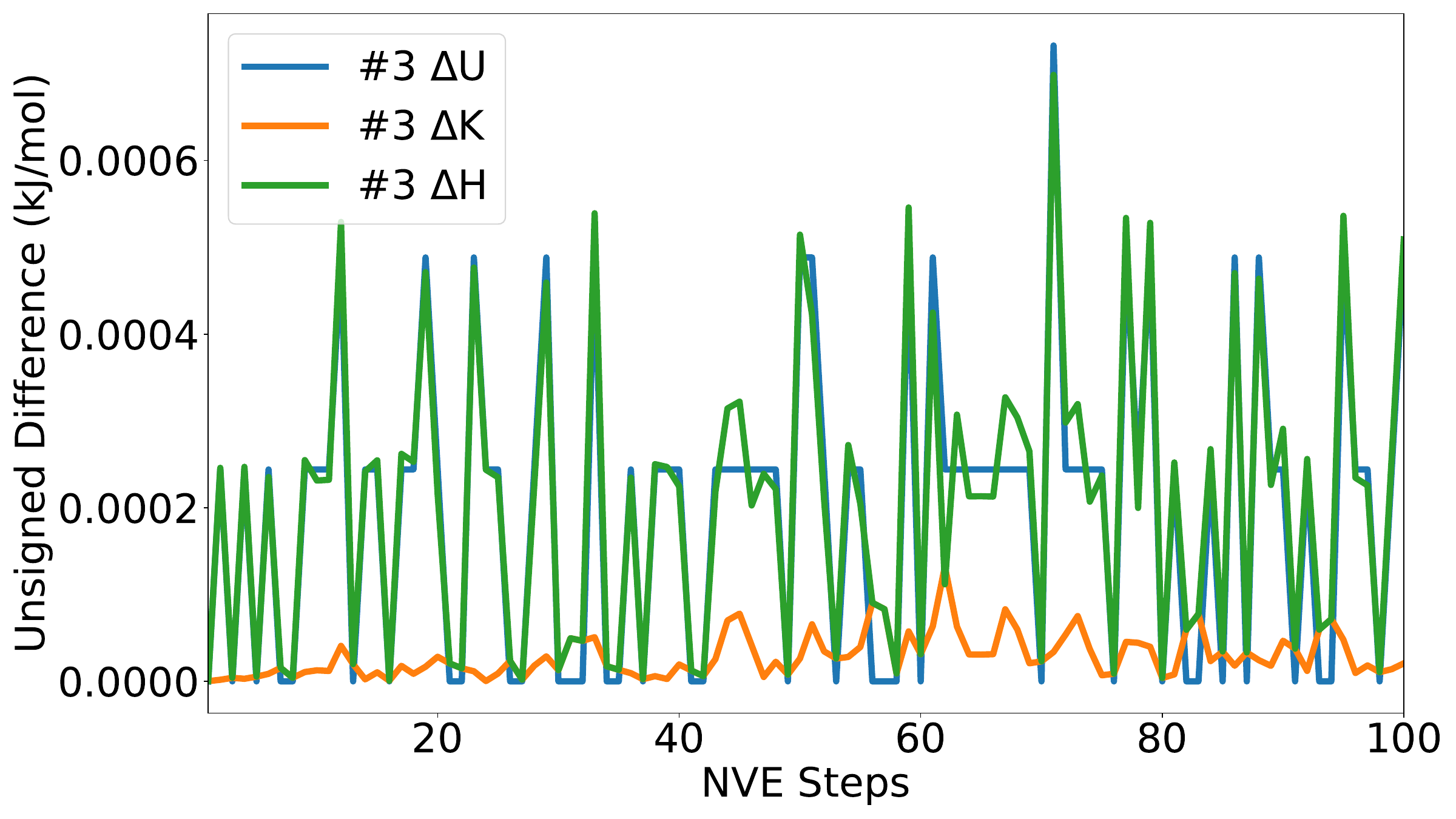}
\end{subfigure}
\begin{subfigure}{0.48\textwidth}
\includegraphics[width=\linewidth]{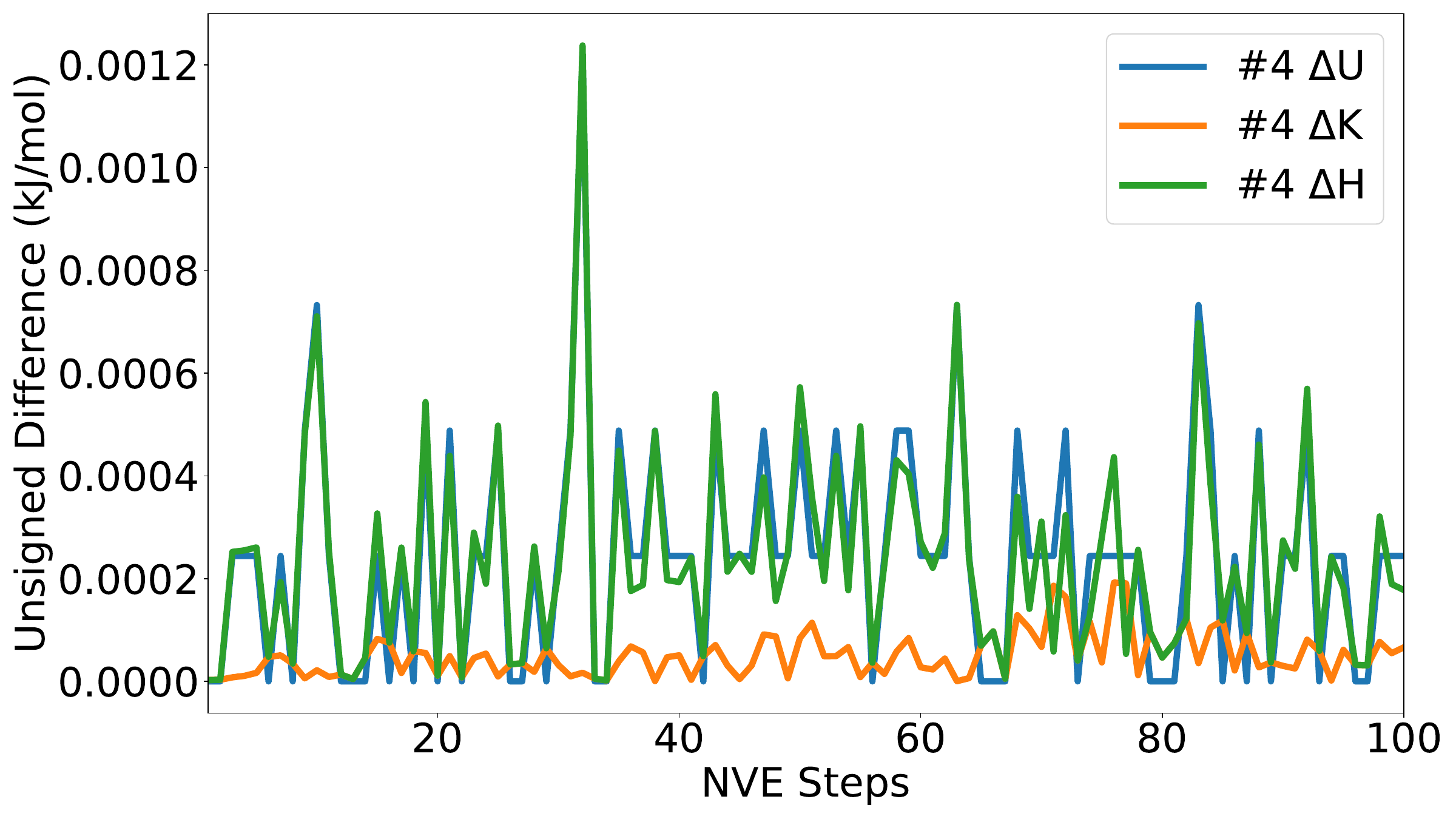}
\end{subfigure}
\\
\begin{subfigure}{0.48\textwidth}
\includegraphics[width=\linewidth]{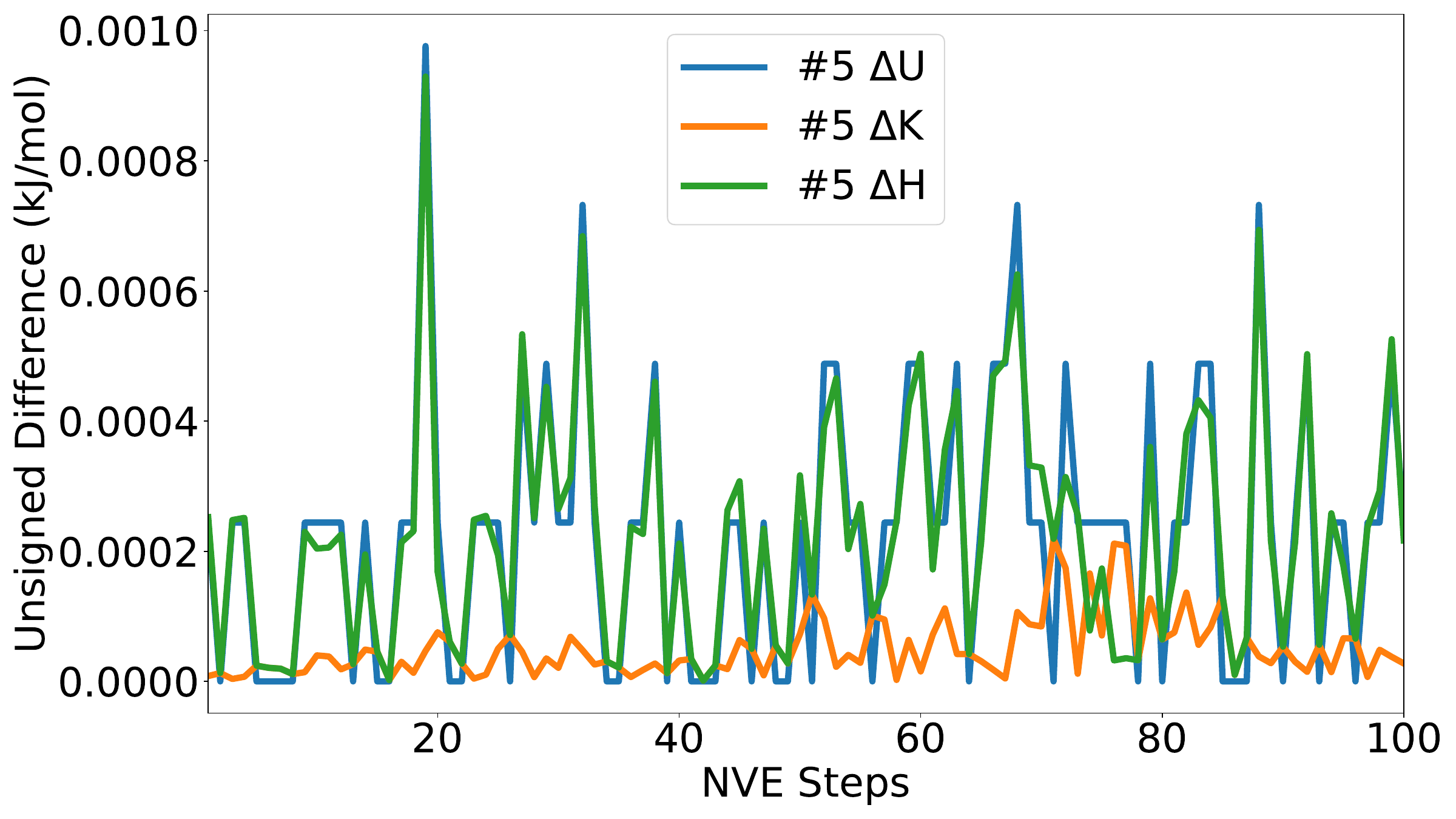}
\end{subfigure}
\begin{subfigure}{0.48\textwidth}
\includegraphics[width=\linewidth]{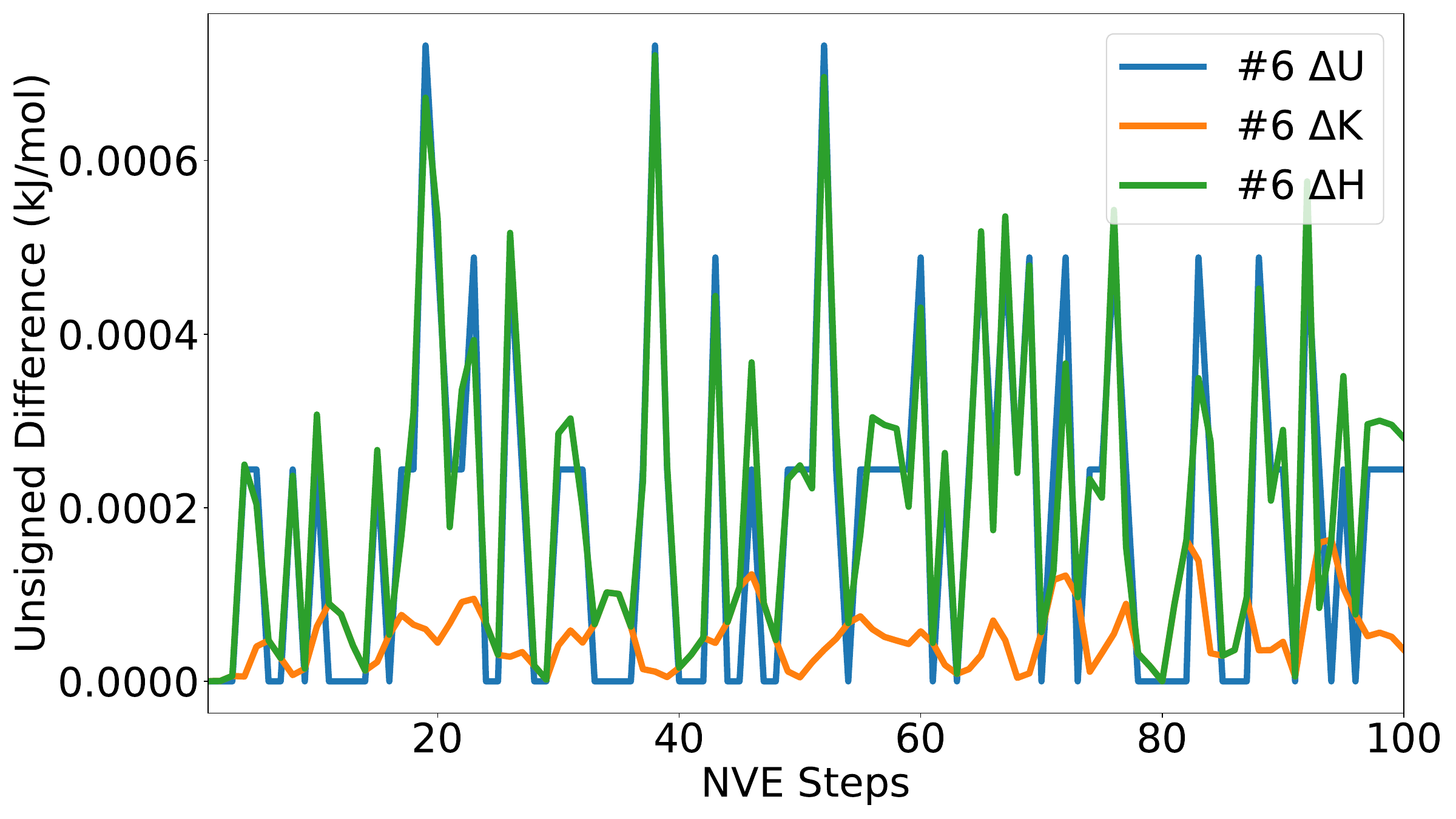}
\end{subfigure}
\\
\begin{subfigure}{0.48\textwidth}
\includegraphics[width=\linewidth]{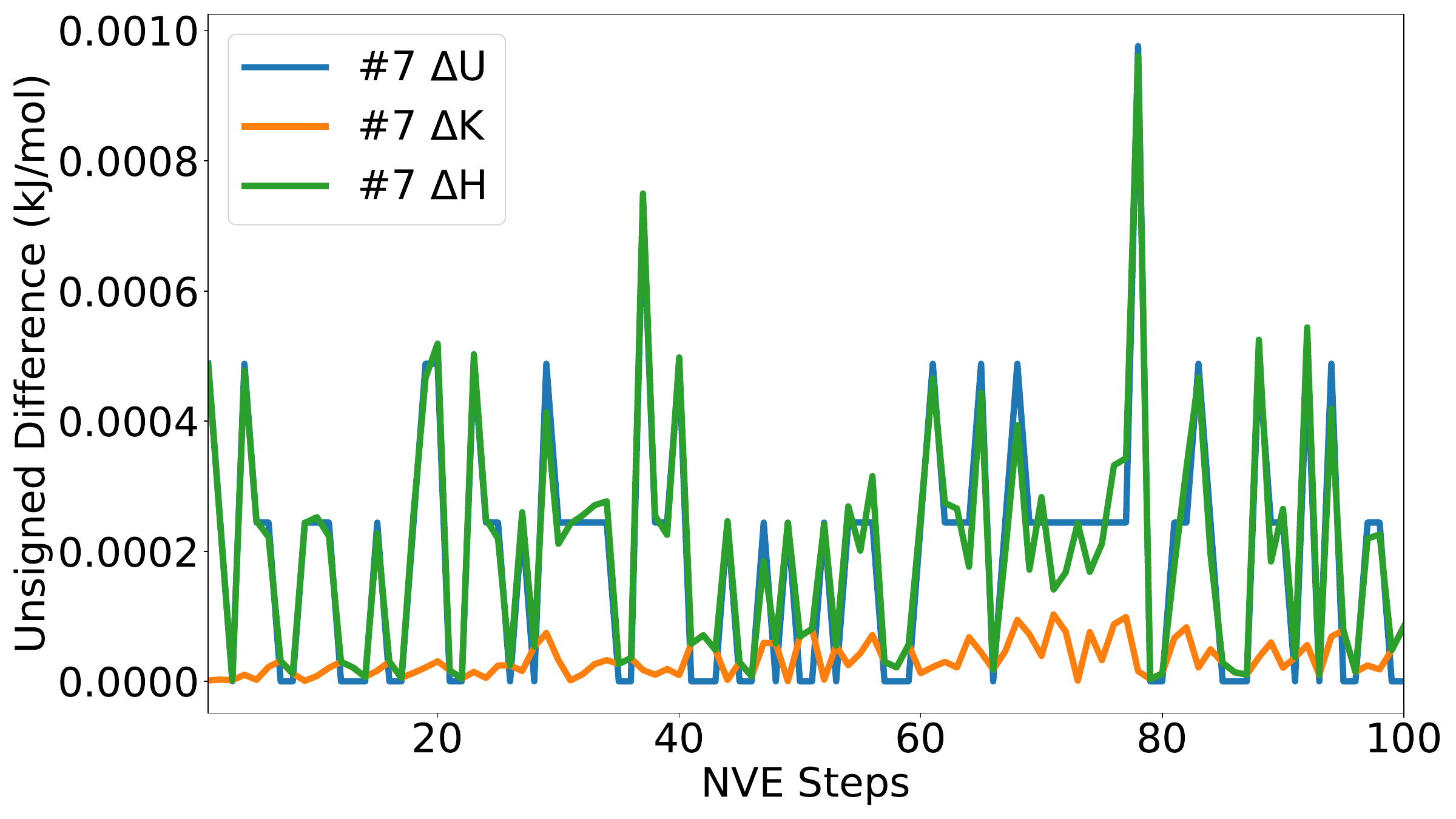}
\end{subfigure}
\begin{subfigure}{0.48\textwidth}
\includegraphics[width=\linewidth]{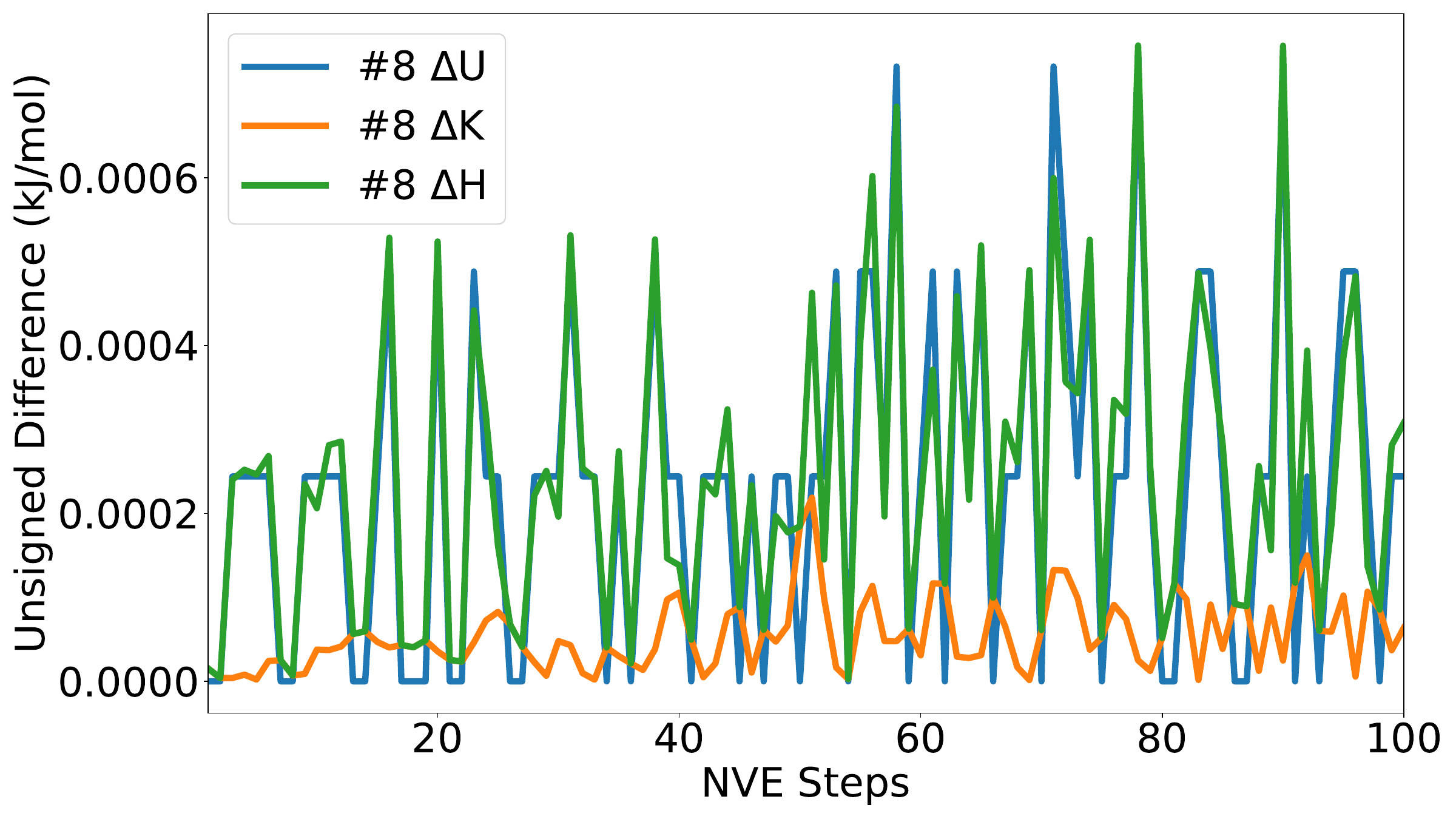}
\end{subfigure}
\caption{Comparison of energies across deployment strategies:
unsigned differences in potential energy ($U$), kinetic energy ($K$), and
Hamiltonian ($H$) relative to the baseline simulation over 100 time-steps.
The number in each subfigure indicates the deployment method
as defined in Table~\ref{tb:deploy8}.}\label{fig:convergence}
\end{figure*}
 
\begin{figure*}[htbp]
\begin{subfigure}{0.48\textwidth}
\includegraphics[width=\linewidth]{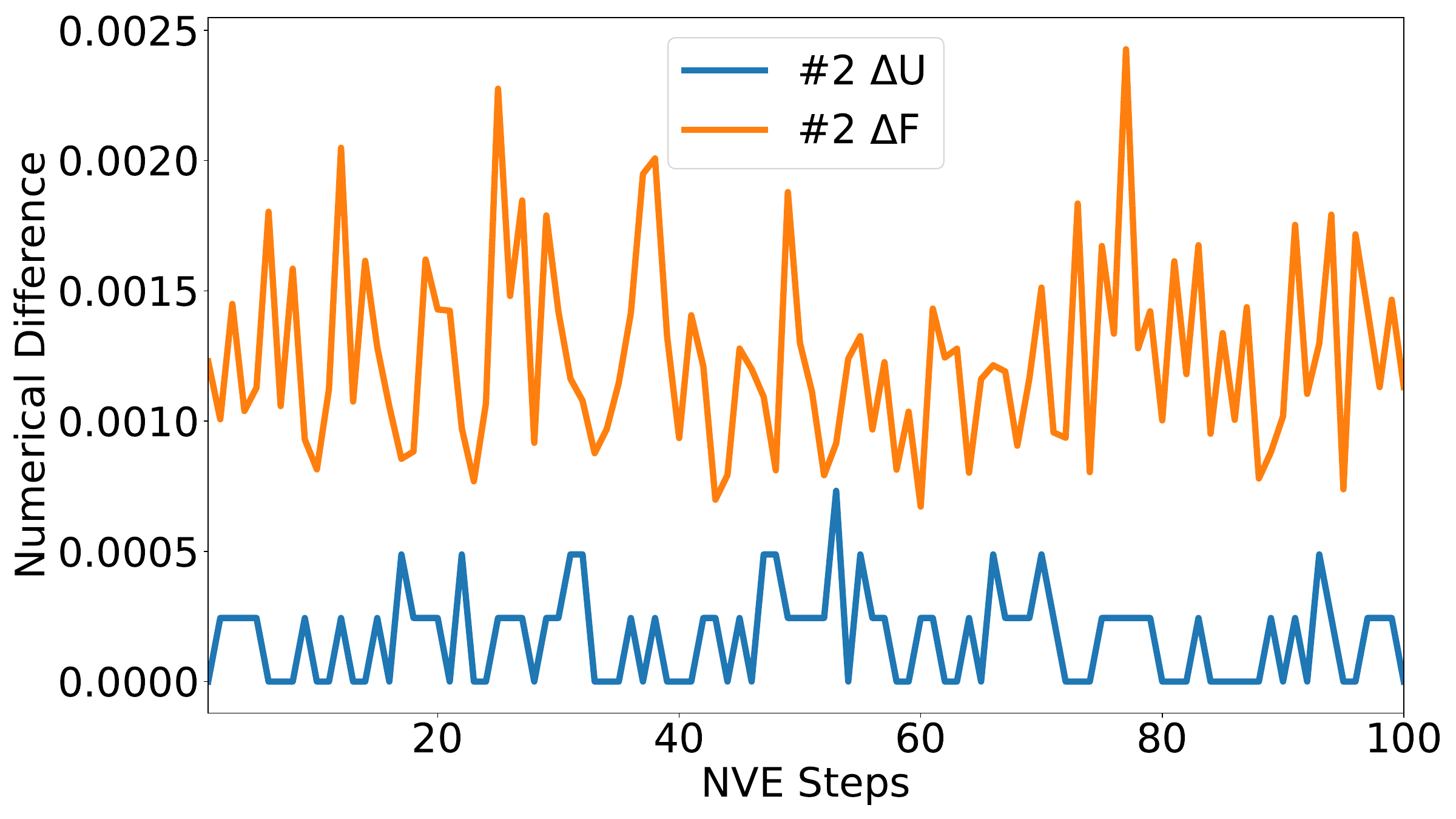}
\end{subfigure}
\\
\begin{subfigure}{0.48\textwidth}
\includegraphics[width=\linewidth]{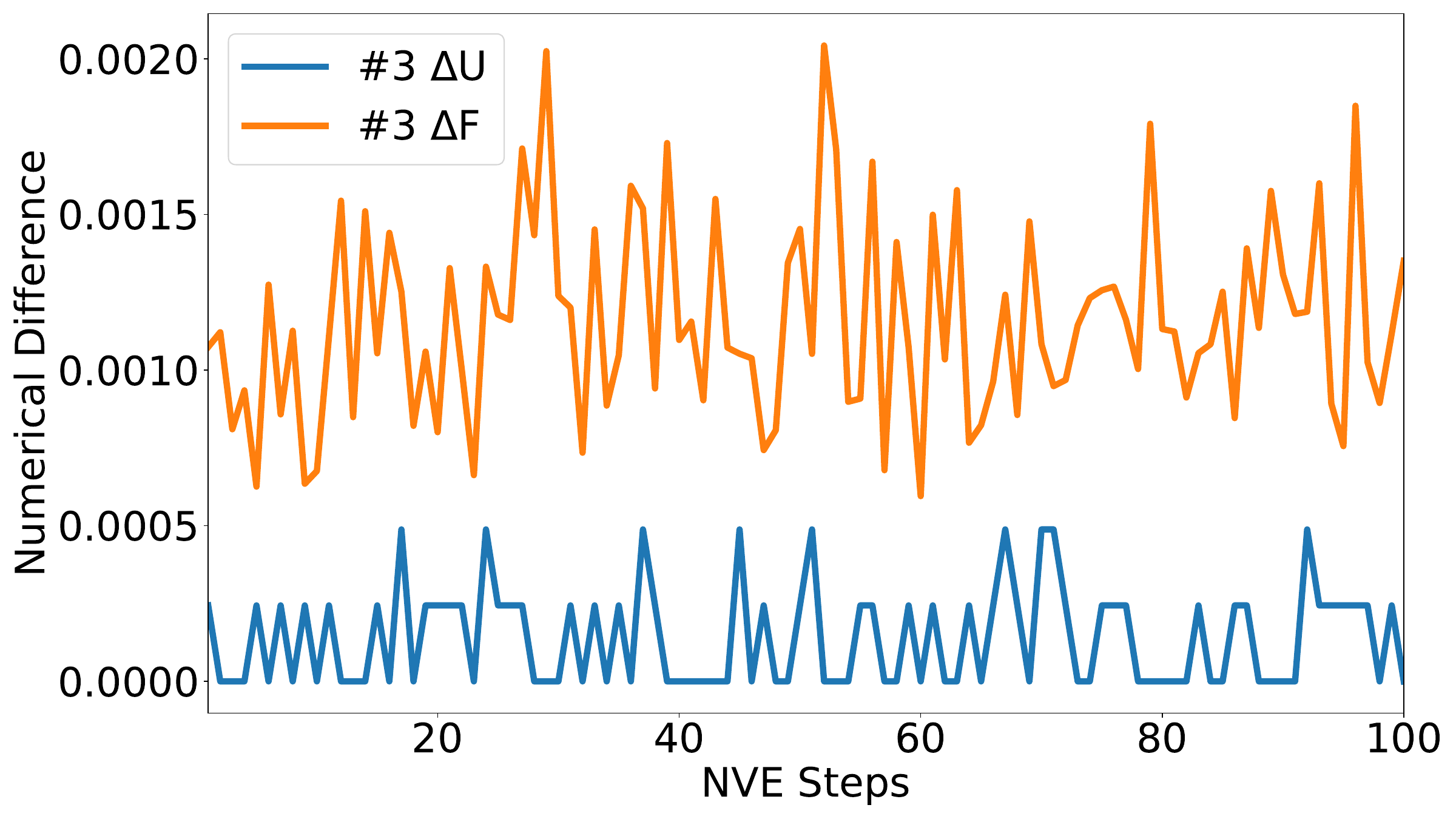}
\end{subfigure}
\begin{subfigure}{0.48\textwidth}
\includegraphics[width=\linewidth]{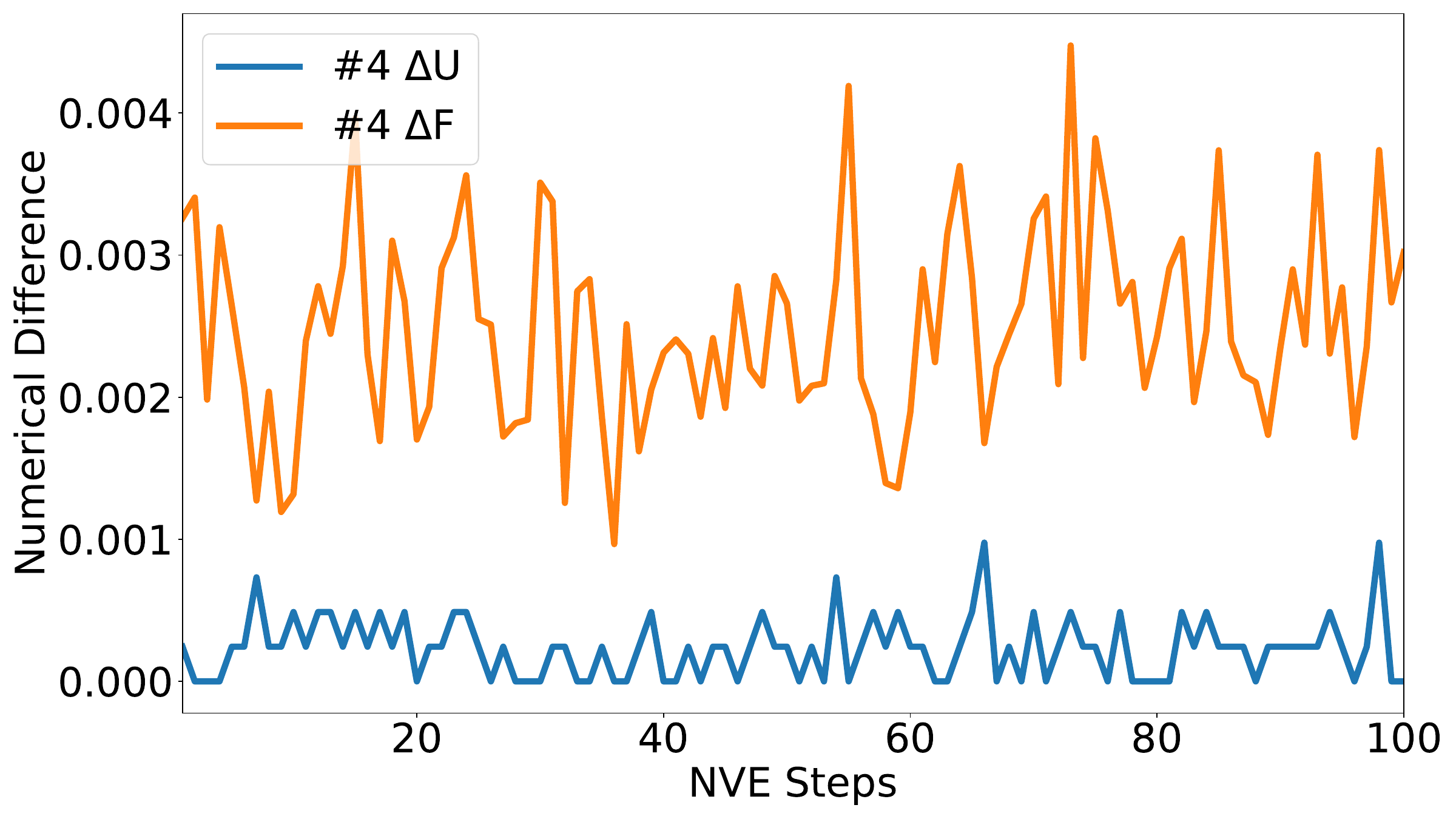}
\end{subfigure}
\\
\begin{subfigure}{0.48\textwidth}
\includegraphics[width=\linewidth]{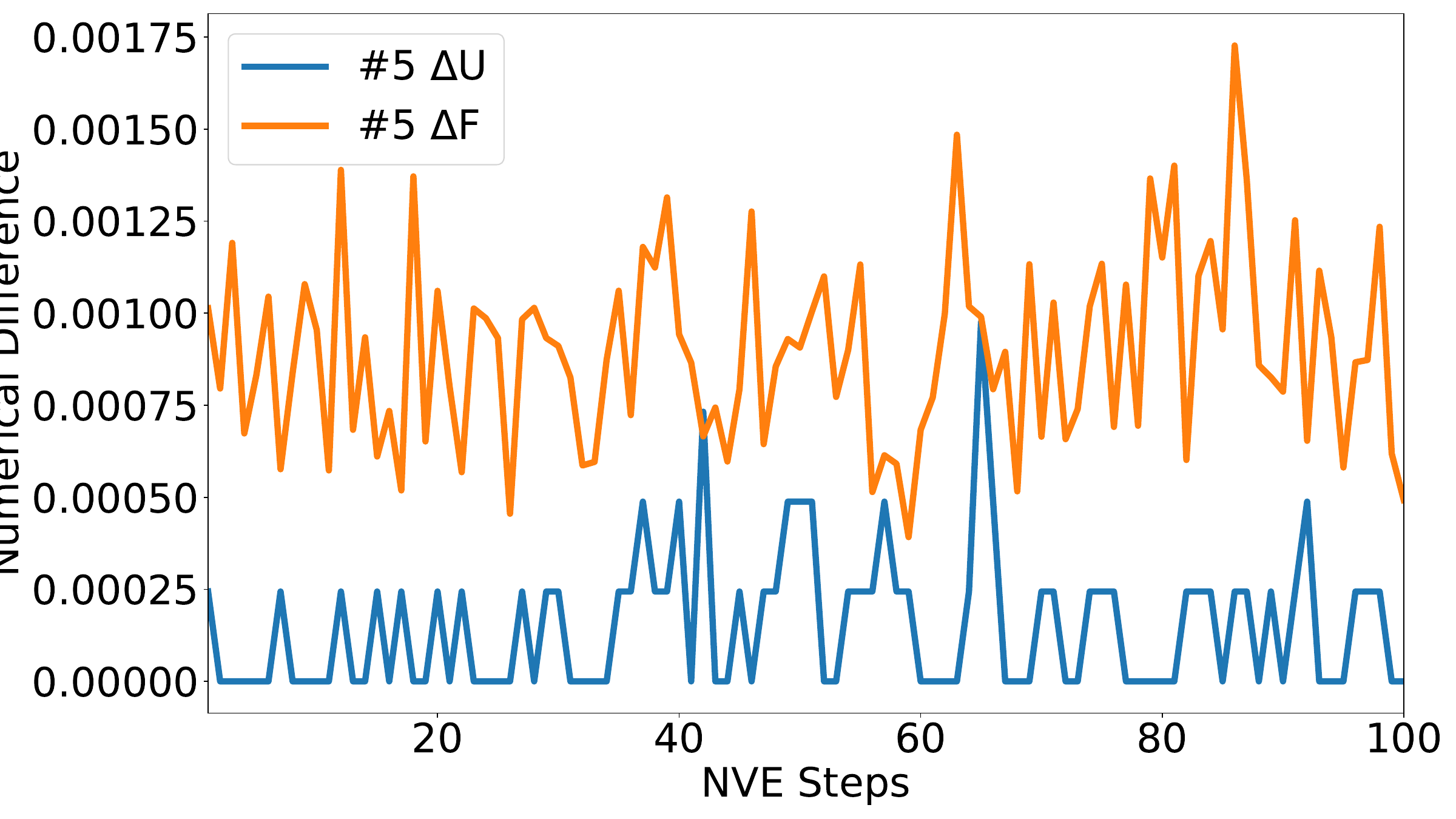}
\end{subfigure}
\begin{subfigure}{0.48\textwidth}
\includegraphics[width=\linewidth]{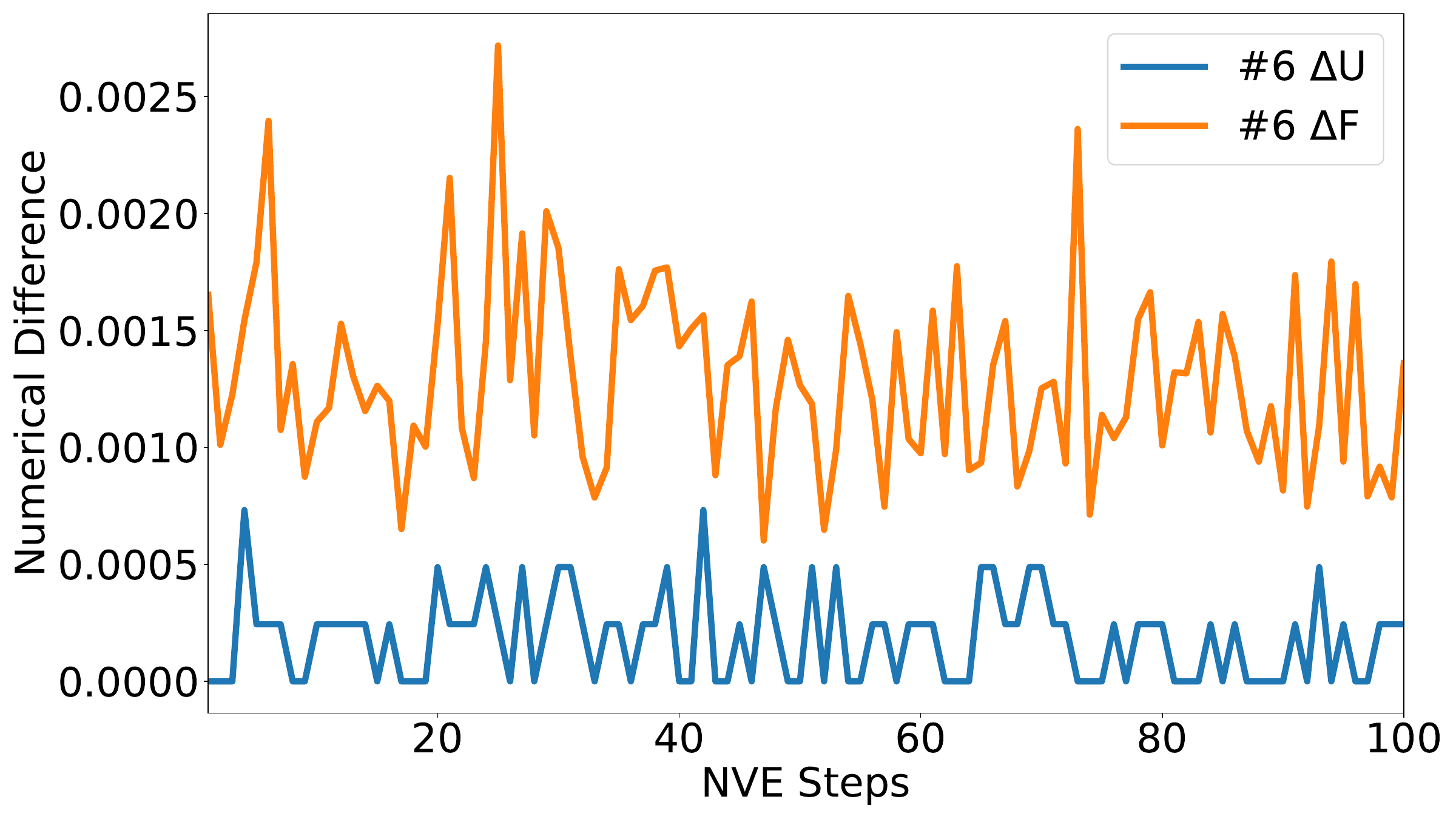}
\end{subfigure}
\\
\begin{subfigure}{0.48\textwidth}
\includegraphics[width=\linewidth]{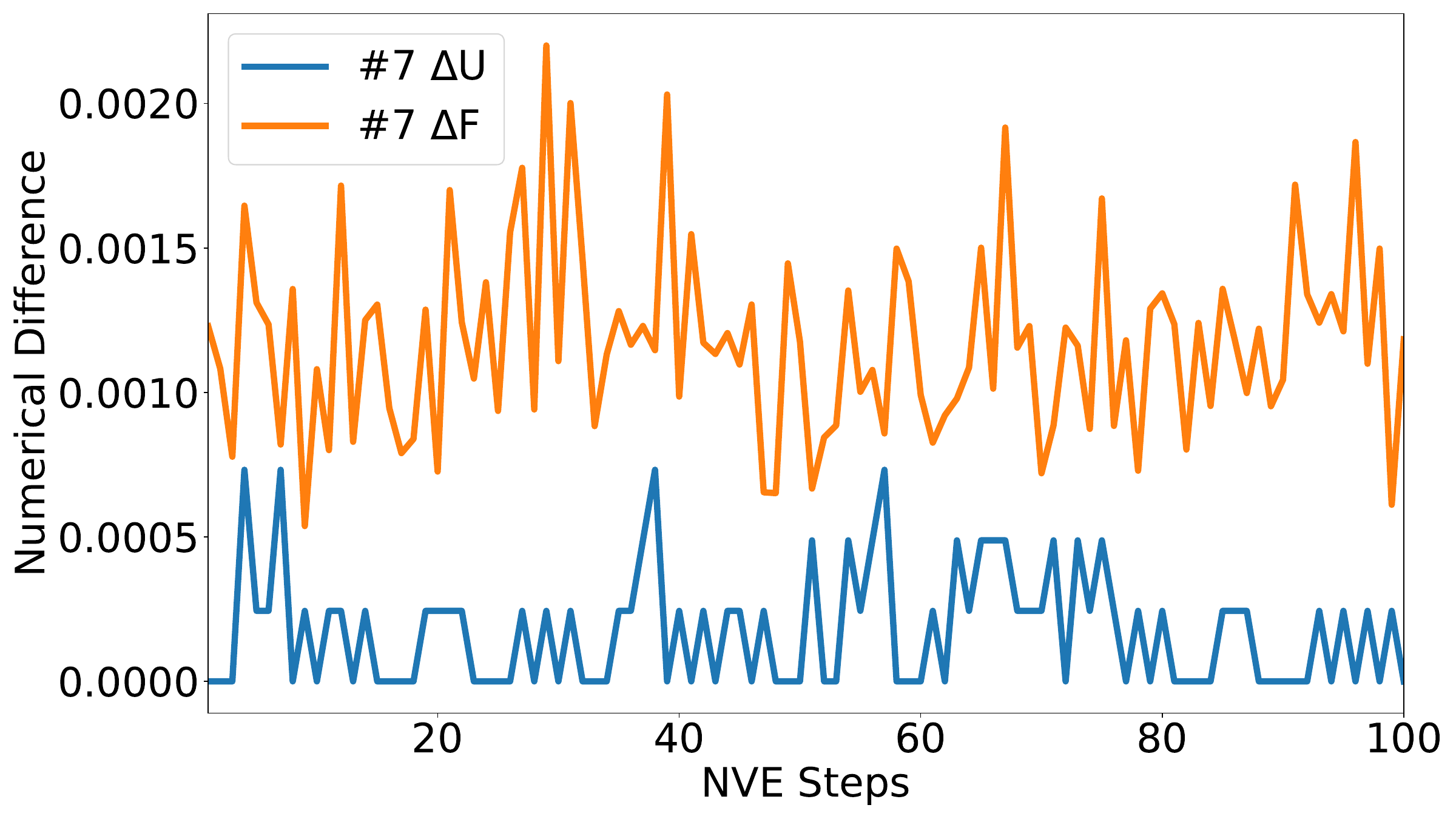}
\end{subfigure}
\begin{subfigure}{0.48\textwidth}
\includegraphics[width=\linewidth]{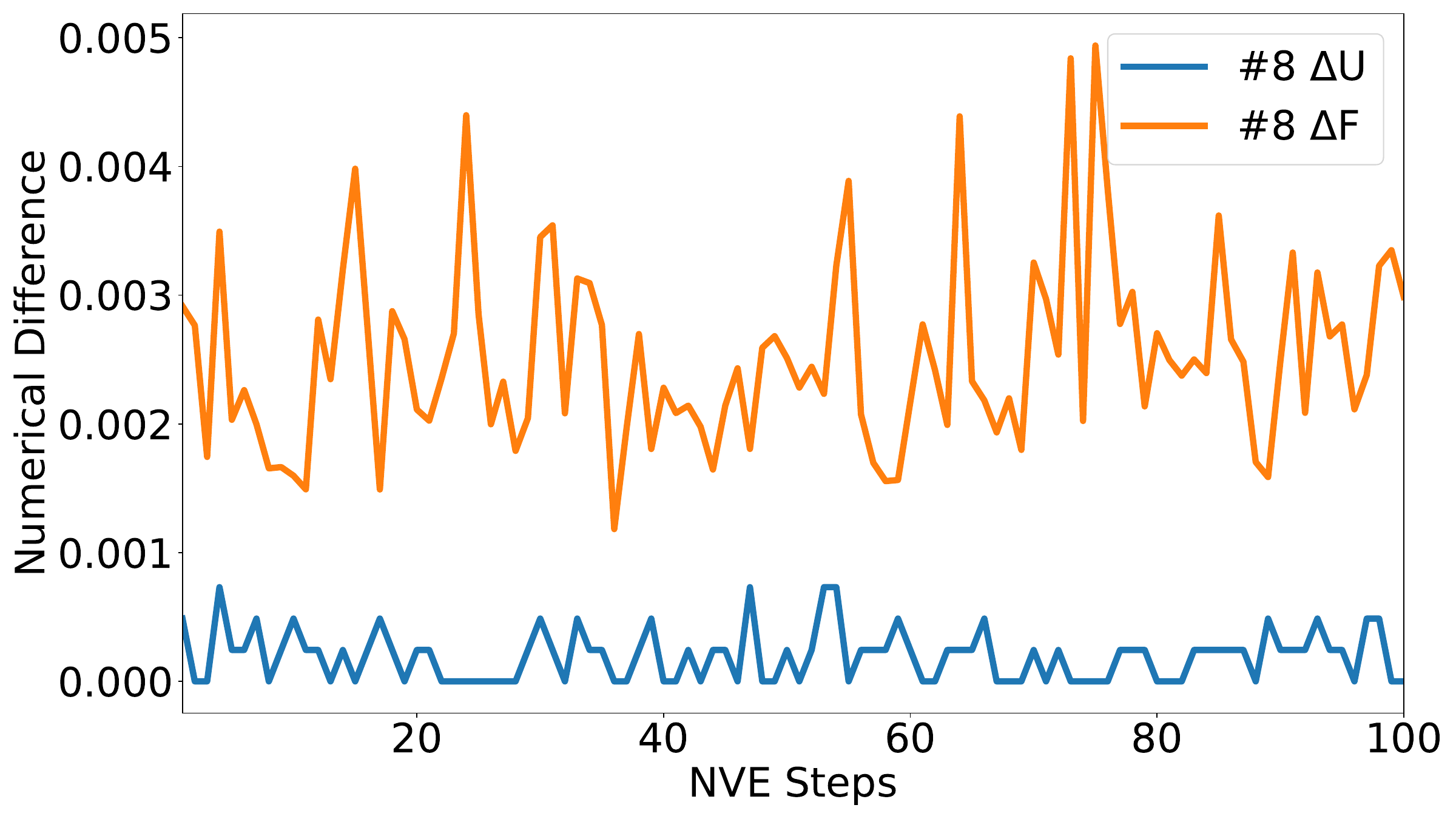}
\end{subfigure}
\caption{The unsigned differences in potential energies
and root mean square deviation of forces (in kJ/mol and nm)
compared to the baseline,
obtained by recalculating an identical PDB trajectory
across deployment methods.
The number in each subfigure denotes the deployment method
as defined in Table~\ref{tb:deploy8}.}
\label{fig:reruns}
\end{figure*}
  \fi

\end{document}